\documentclass[final,5p,times,twocolumn]{elsarticle}

\usepackage{graphicx}
\usepackage{amssymb}
\usepackage{amsmath}
\usepackage{empheq} 
\usepackage{comment}
\usepackage{balance}
\usepackage{enumerate}
\usepackage[inline,shortlabels]{enumitem}
\usepackage[usenames,dvipsnames]{xcolor}
\usepackage[colorlinks=true,linkcolor=Maroon,citecolor=Maroon,urlcolor=Maroon]{hyperref} 
\usepackage{url}
\usepackage{floatpag}
\usepackage{footmisc}
\usepackage{booktabs} 
\usepackage{balance} 
\usepackage{subcaption}
\usepackage{caption}
\usepackage[normalem]{ulem}
\usepackage{threeparttable}
\usepackage{multirow}

\biboptions{sort&compress} 

\makeatletter
\def\ps@pprintTitle{%
 \let\@oddhead\@empty 
 \let\@evenhead\@empty
 \def\@oddfoot{}%
 \let\@evenfoot\@oddfoot}
 \makeatother

\begin{document}
\begin{frontmatter}

\title{
Data-Driven Law Firm Rankings to Reduce Information Asymmetry in Legal Disputes}

\author[add1]{Alexandre Mojon}
\author[add2,add3]{Robert Mahari}
\author[add2,add4]{Sandro Claudio Lera$^\dagger$}

\address[add1]{\scriptsize University of St. Gallen, St. Gallen, Switzerland}
\address[add2]{\scriptsize Massachusetts Institute of Technology, Cambridge, USA}
\address[add3]{\scriptsize Harvard Law School, Cambridge, USA}
\address[add4]{\scriptsize Southern University of Science and Technology, Shenzhen, China}
	
\begin{abstract}
Selecting capable counsel can shape the outcome of litigation, yet evaluating law firm performance remains challenging. Widely used rankings prioritize prestige, size, and revenue rather than empirical litigation outcomes, offering little practical guidance. To address this gap, we build on the Bradley-Terry model and introduce a new ranking framework that treats each lawsuit as a competitive game between plaintiff and defendant law firms. Leveraging a newly constructed dataset of 60,540 U.S. civil lawsuits involving 54,541 law firms, our findings show that existing reputation-based rankings correlate poorly with actual litigation success, whereas our outcome-based ranking substantially improves predictive accuracy. These findings establish a foundation for more transparent, data-driven assessments of legal performance.  

\end{abstract}
\end{frontmatter}

\renewcommand*{\thefootnote}{\fnsymbol{footnote}} 
\footnotetext[0]{$^\dagger$Corresponding author: \texttt{slera@mit.edu}. \\The authors declare no competing interest.}

\section*{Introduction}

Legal representation plays a crucial role in determining litigation outcomes.
Research consistently shows that litigants represented by attorneys fare better in court than those without legal representation~\cite{poppe2015lawyers}.
In addition, the quality of representation significantly affects the outcomes of the cases.
In asylum proceedings, for example, empirical evidence suggests that individuals with better attorneys have a higher probability of prevailing~\cite{Miller2015}.
This effect extends beyond individual cases, as disparities in legal expertise shape broader patterns of judicial decision-making~\cite{Szmer2007}.

Despite the importance of legal representation, most litigants lack the necessary information to identify the most effective attorney for their case.
Legal expertise is highly specialized, and a firm’s ability to succeed in one area of law does not necessarily translate to success in another.
Adding to this challenge, the information asymmetry in the legal market favors those with more financial resources, who can afford to seek specialized advice or hire firms with established reputations~\cite{Sandefur2015,poppe2015lawyers}.
Previous research suggests that this knowledge gap reinforces structural inequalities, enabling well-resourced litigants to gain advantages independent of the merits of the case~\cite{Galanter1974,Rhode2004,Garrett2011,Cornwell2017}.

The most widely used law firm rankings do little to address this problem.
Existing rankings, such as Vault 100 and ALM’s Global 200, emphasize prestige, firm size, or revenue rather than empirical litigation performance.
Furthermore, these rankings usually cover only the top 100 or top 300 firms, excluding a vast majority of firms that may be more suitable for specific cases.
Factors such as prestige and firm size do not necessarily reflect the ability of a firm to achieve favorable outcomes for its clients.
Reputation-based marketing can perpetuate a feedback loop, in which high-ranking firms attract more clients, reinforcing their prestige and market dominance, regardless of their actual effectiveness in court~\cite{Zacharias2008}.
Furthermore, such rankings are not tailored to specific case types or areas of legal expertise, making them of limited use to litigants seeking the right lawyer for their case.

\begin{figure*}[!htb]
    \centering
    \includegraphics[width=\linewidth]{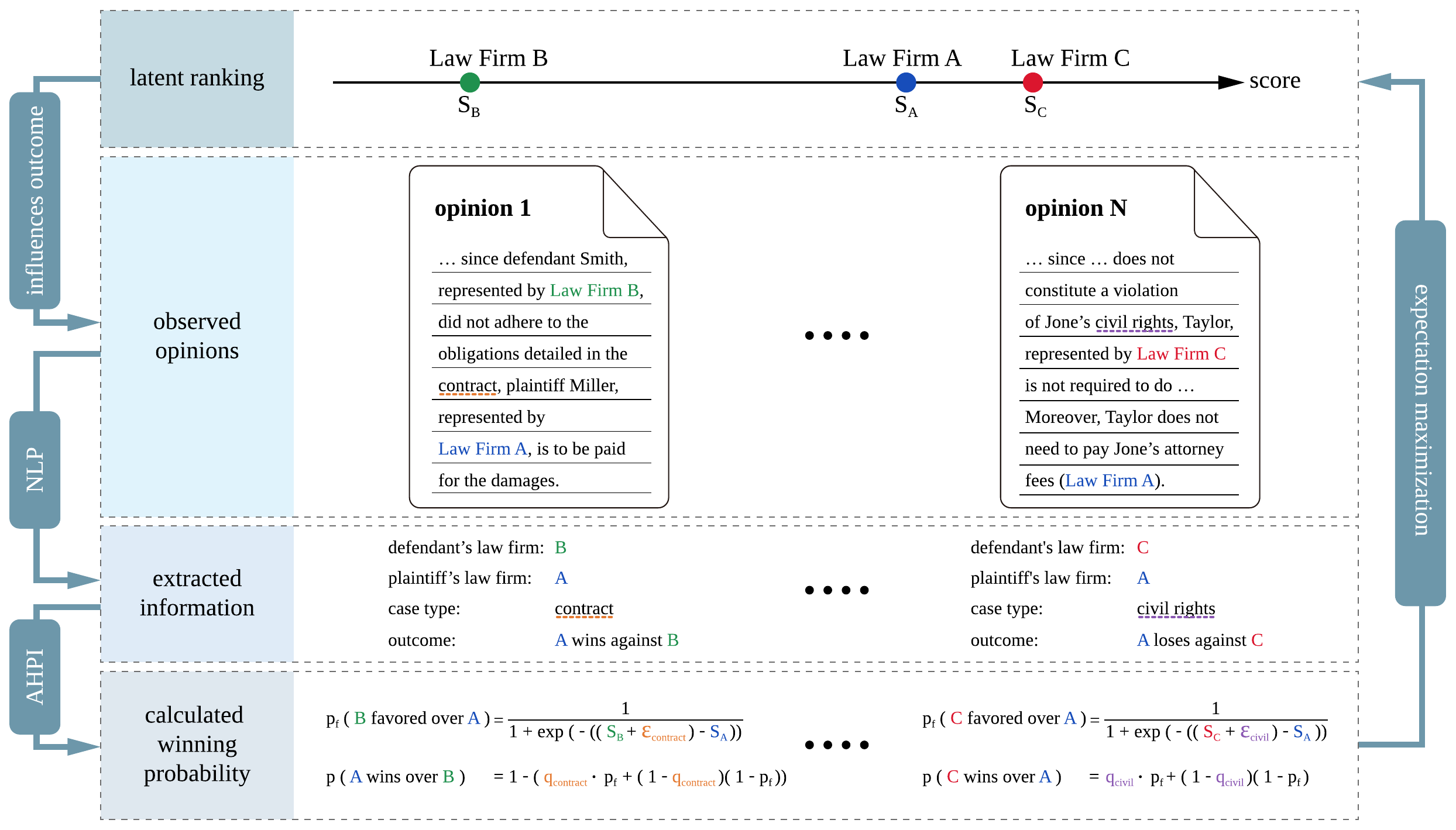}
    \caption{
	We analyze $N$ lawsuits based on judges' textual decisions (\textit{opinions}),  
	from which we extract structured information on the case type, the case outcome, and the law firms involved.  
	Each lawsuit is modeled as a \textit{game} between the plaintiff's and defendant's law firm, where one firm wins and the other loses.  
	Each of the five case types has an associated defendant bias (\textit{home field advantage}) $\epsilon_m$,  
	which quantifies the a priori likelihood that the defendant prevails in case type $m$.  
	Each law firm $k = 1, \ldots, K$ is assigned a latent skill score, $S_k$.  
	The propensity that defendant's law firm $B$ is \textit{favored} over the plaintiff's law firm $A$ is modeled as a sigmoid function of $(S_B + \epsilon_m) - S_A$,  
	such that a higher bias-adjusted score for $B$ increases its likelihood of being favored.  
	The valence probability $q_m$ then determines the probability that the favored firm ultimately wins, capturing the role of uncertainty in case outcomes.
	Using these assumptions, we apply an expectation-maximization algorithm to infer the latent law firm scores $\{ S_k \}$,  
	defendant biases $\{\epsilon_m\}$, 
	and valence probabilities $\{q_m\}$  
	that best explain the observed litigation outcomes.  
	These rankings are then used to predict case outcomes, outperforming existing reputation-based rankings (Figure \ref{fig:rank_clock}).  
	The inferred defendant biases $\epsilon_m$ are consistently positive (Table \ref{tab:case_summary}),  
	reflecting the well-documented advantage of defendants in litigation.  
	The valence probabilities $q_m$ are found to be high (between 85\% and 100\%, as shown in Table \ref{tab:case_summary}), underscoring the significant role law firms play in shaping case outcomes.
    }
    \label{fig:sketch}
\end{figure*}

To address these limitations, we introduce a methodology for ranking law firms based on objective case outcomes rather than subjective prestige metrics.  
To this end, we generalize a ranking algorithm designed for heterogeneous pairwise zero-sum games \cite{Newman2022}, where \textit{games} represent lawsuits and \textit{players} are law firms.  
This model, based on the Bradley-Terry framework \cite{BradleyTerry1952}, ranks entities by their probability of winning based on their relative strengths or \textit{scores}.
To enable a more accurate ranking of law firms across various legal domains, we account for different case types and asymmetric outcomes, namely that defendants are a priori more likely to win (akin to a \textit{home field advantage})~\cite{eisenberg1995litigation}.
Given a set of observed lawsuits (game outcomes), we fit an expectation maximization algorithm to estimate the law firms' latent scores, which are then interpreted as law firm rankings.  
Once fitted, we can predict the outcome of a lawsuit as a function of the difference between the scores of the two law firms: the higher the score of law firm A over law firm B, the more likely A is to win (see Figure \ref{fig:sketch} for an overview). 

To test our ranking methodology, we compile a comprehensive dataset that amalgamates several preexisting data sources and extracts pertinent information from judges' published decisions.  
This dataset encompasses detailed information of 60,540 civil lawsuits in U.S. Federal District Courts, including the names of the law firms representing plaintiffs and defendants, respectively, alongside the case outcome, that is, whether the plaintiff or defendant won the case.  
Fitting this data with our methodology yields law firm scores for 2,064 law firms.
We find that these scores do not have a significant correlation with the prestige-based law firm rankings, but significantly outperform them in predicting future lawsuit outcomes.

In summary, we introduce an outcome-based law firm ranking system that outperforms prestige-focused rankings in predicting case outcomes, challenging traditional metrics, and promoting a more transparent legal system.
By prioritizing effectiveness over reputation, our approach provides a more practical assessment of law firm quality, helping litigants make informed choices.

\section*{Results}

\subsection*{A Dataset of Zero-Sum Law Games}

We ground our ranking of law firms in a novel dataset of 60,540 civil lawsuits in U.S. federal courts. 
This dataset distinguishes itself from previous work by focusing not only on the presence or absence of legal representation, but also including the identity of law firms~\cite{Mahari2024}.
In addition, while most previous studies focus on disputes involving individuals and relatively small amounts in controversy such as
small claims court~\cite{Lafree1996}, 
family law~\cite{Maccoby1992}, 
and housing law~\cite{Hannaford2003},
our dataset covers a wide range of disputes from civil rights and torts to large commercial disputes.

Our dataset is constructed from judicial opinions provided by the Case Law Access Project (CAP), which contains textual records of U.S. Federal District Court opinions.  
These opinions include detailed case descriptions written by judges but do not directly provide structured information on case types, law firms, or outcomes.  
To extract this information, we leverage a fine-tuned transformer language model to classify case types and determine case outcomes based on the judge’s opinion (see Methods).  
Five major case types are distinguished---civil rights, contracts, labor, torts, and \textit{other}---which vary in their observed frequencies (Table~\ref{tab:case_summary}).  
The case outcome is defined as a binary variable, where a value of 1 indicates a plaintiff victory and 0 indicates a defendant victory.  
To identify the law firms involved, we extract firm names from the judge’s opinion using a pattern-matching approach and cluster similar names to account for variations in spelling and formatting (see Methods). 
This process yields a dataset comprising 60,540 civil cases involving 54,541 law firms. 
If a case involves multiple law firms representing either the plaintiff or the defendant, 
we decompose it into pairwise interactions, each consisting of one plaintiff and one defendant law firm 
(see \textit{Limitations} for potential generalizations of this assumption).  
This results in a total of 190,297 pairwise interactions across 54,541 law firms.
For each interaction, we have structured data including the names of the plaintiff and defendant law firms, the case type, and the case outcome (either a plaintiff or defendant victory).  
We use the first 80\% of cases to fit the law firm scores and reserve the remaining 20\% to evaluate the predictive accuracy of the resulting rankings.  

Our training dataset allows us to construct an \textit{interaction network} of law firms, where each pairwise interaction between a plaintiff’s and a defendant’s law firm is represented as an undirected edge.
This interaction network exhibits considerable variation in density, as some law firms participate in many cases while others appear only a few times.  
To systematically analyze this structure, we introduce the \textit{$Q$-factor}, which quantifies the average number of observed interactions per law firm within a given subset of the network.  
We trim the network by iteratively removing the law firms with lowest number of interactions until the remaining network meets a specified $Q$-factor threshold.  
Setting $Q$ sufficiently high ensures that each law firm has a meaningful number of observed interactions, allowing us to reliably fit its score.  
In the main article, we use $Q=30$ to compute law firm rankings, resulting in an interaction network with $N=63,115$ interactions involving $K=2,064$ law firms.  
We verify in the SI Appendix that our results remain qualitatively similar across a range of $Q$ values from $20$ to $55$.  

While our dataset provides an empirical foundation for ranking law firms based on case outcomes, it is important to acknowledge its limitations.  
Most notably, it includes only cases that resulted in a published judicial opinion, excluding those settled out of court.  
Since the majority of legal disputes are resolved through private settlements~\cite{prescott2016comprehensive}, 
our dataset captures only a selective subset of legal interactions---specifically, those where litigants proceeded to trial.
We discuss this limitation and potential extensions incorporating settled cases below.  
Despite these constraints, our methodology remains robust under different data filtering approaches and offers a computational framework for ranking law firms using objective litigation outcomes.

\begin{table}[ht]
\centering
\begin{tabular}{l|ccccc}
\hline
\textbf{Type} & \textbf{Share} & \textbf{Win Rate} & \textbf{Spec.} & $\hat{\boldsymbol{\epsilon}}$ & $\hat{\mathbf{q}}$ \\ \hline\hline
Civil Rights & 22.9\%  & 86.0\% & 16.1\% & 2.03 & 0.86 \\
Contract     & 18.6\%  & 70.4\% & 14.4\% & 1.66 & 0.96 \\
Torts        & 13.2\%  & 85.0\% & 12.4\% & 0.32 & 1.00 \\
Labor        & 8.8\%   & 73.7\% & 6.2\%  & 1.99 & 0.96 \\
Other        & 36.4\%  & 75.2\% & 28.0\% & 1.90 & 1.00 \\ \hline\hline
\end{tabular}
\caption{
Summary statistics calculated across 60,540  civil cases.
(\textbf{Share})
Total fraction of cases corresponding to each case type.
(\textbf{Win Rate})
Average defendant win rate per case type.
(\textbf{Spec.})
Fraction of law firms which specialize in only this case type.
(\textbf{$\hat{\boldsymbol{\epsilon}}$})
Defendant biases fitted from the data. 
(\textbf{$\hat{\mathbf{q}}$})
Valence probabilities fitted from the data.
}
\label{tab:case_summary}
\end{table}

\subsection*{Ranking Law Firms via Asymmetric Pairwise Interactions}

\begin{figure}[!htb]
    \centering
    \includegraphics[width=\linewidth]{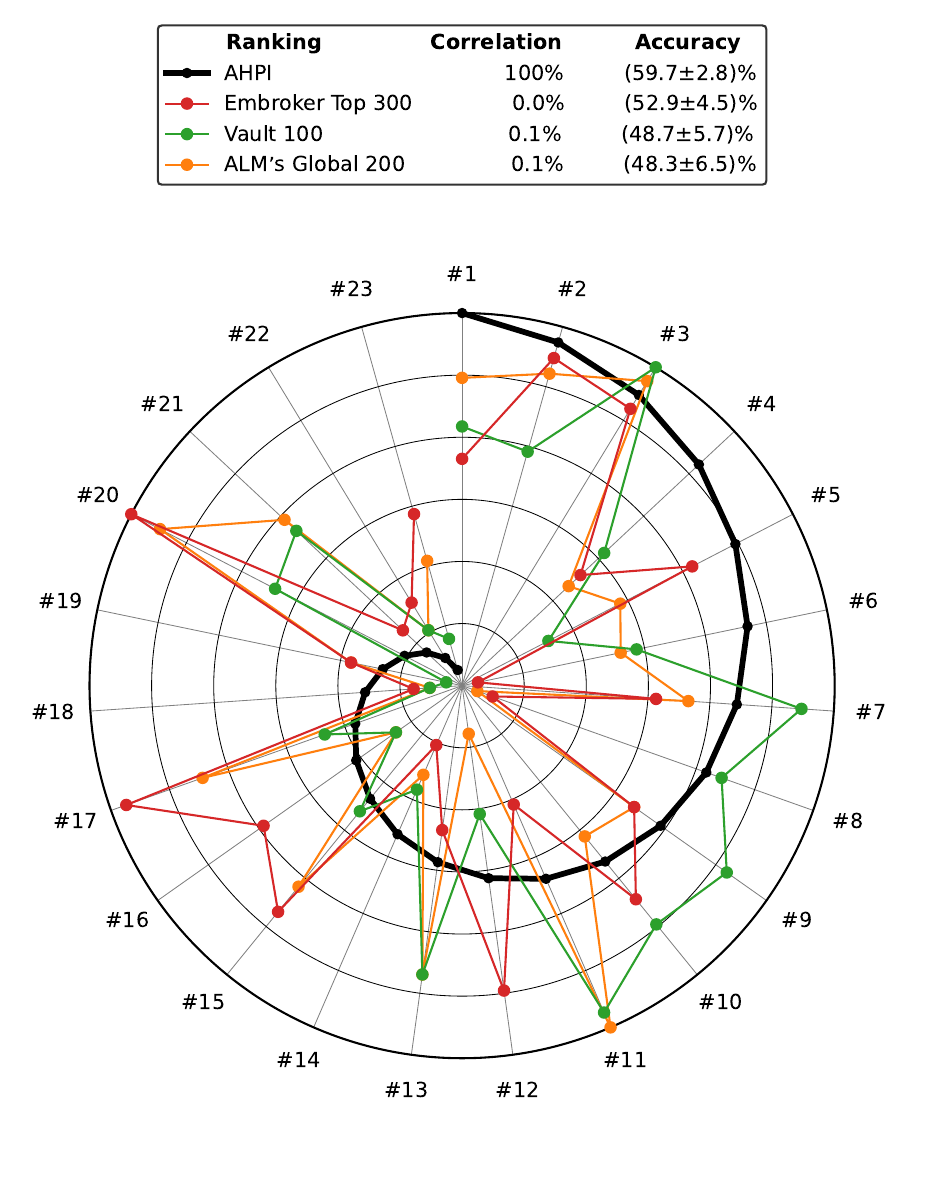}
    \caption{
    We compare our AHPI scores with three widely used firm rankings: 
    \textit{Vault 100, ALM's Global 200} and \textit{Embroker Top 300}.
    (top)
    Correlation between AHPI ranking and each of the these rankings, 
    and predictive accuracy on test cases given each ranking, 
    where  50\% represents the expected accuracy from random guessing. 
    (bottom)
    Comparison of ranks for 20 law firms which are common across all four rankings.
    A point close to the periphery means a higher rank than one closer to the center.
    As visually apparent and indicated by the low correlation coefficients in the `Correlation' column, 
    there are significant differences between AHPI scores and other rankings.
    }
    \label{fig:rank_clock}
\end{figure}

We use the above dataset, consisting of $N=63,115$ interactions involving $K=2,064$ law firms (corresponding to $Q=30$), to develop an objective metric of the performance of law firms. 
A naive approach would be to rank firms based on their overall win rate.  
However, this method has two key limitations:
First, there is an intrinsic bias toward defendant victories (analogous to a \textit{home field advantage} in sports)~\cite{eisenberg1995litigation,Lera2022,Mahari2024}.
Consequently, a plaintiff’s win is not an equivalent achievement to a defendant’s, necessitating an adjustment for this asymmetry. 
Second, different case types (\textit{rules of the game}) exhibit varying baseline win rates for defendants (Table~\ref{tab:case_summary}).  
Although law firms specialize to some extent, they generally engage in multiple case types, meaning that wins and losses occur under different conditions.
These challenges are further compounded by the presence of noise or \textit{luck} in litigation outcomes, where the superior firms may lose and the weaker firms may win \cite{Jerdee2023}.  
We therefore propose a ranking method that accounts for these structural biases.

Consider a scenario in which $K$ entities engage in $N$ pairwise interactions (\textit{games}) with a winner and a loser, such as in sports tournaments or, in our case, lawsuits.  
The objective is to assign scores to entities that measure their skill.
A well-known approach to this problem is the probabilistic Bradley-Terry model, which assumes that the probability of entity A with a score $S_A$ defeating entity B with a score $S_B$ is given by a sigmoid function of the difference in their scores $S_A - S_B$~\cite{BradleyTerry1952}.
This model has been extended to incorporate different interaction types~\cite{Newman2022}, a crucial feature for our application where case types significantly influence litigation outcomes. 
A key distinction between our problem and typical ranking models is that litigation outcomes are structurally asymmetric---defendants have a systematically higher likelihood of winning, and this likelihood varies across case types (Table~\ref{tab:case_summary}).  
We thus extend ranking algorithms to account for multiple interaction types (here case types) and systemic defendant advantage.  
We refer to this method as the \textit{Asymmetric Heterogeneous Pairwise Interactions (AHPI)} ranking algorithm.  

The AHPI ranking algorithm employs a Bayesian expectation-maximization framework with a logistic prior over entity scores (see Methods).  
It takes as input $N$ observed interactions with binary outcomes, modeling each as an asymmetric game played between pairs of $K$ entities.  
There is a total of $M$ case types.
The game is asymmetric in the sense that, when the plaintiff's law firm $A$ and the defendant's law firm $B$ have the same estimated score ($S_A=S_B$), the defendant's firm (\textit{home team}) has a higher a priori chance of winning.
Each case type is characterized by a different baseline probability of defendant victory, reflecting systemic legal biases.
The AHPI ranking algorithm operates in two stages: 
First, it estimates the propensity that law firm $A$ is \textit{favored} over law firm $B$ using a sigmoid function of their score difference: 
$S_A - (S_B+\epsilon)$.
Here, the $\epsilon > 0$ represents the bias towards the defendant. 
We assume that each case type $m$ has its own $\epsilon_m$ parameter, and fit it from the data. 
Second, a case-type-specific valence probability $q_m \in [0.5, 1]$ determines whether the \textit{favored} firm actually wins, accounting for the inherent uncertainty in litigation outcomes.
This two-step process provides a natural way to quantify the role of luck in the game \cite{Jerdee2023}, with lower $q_m$ values indicating higher randomness in outcomes.
The $K$ law firm scores, the $M$ case-specific defendant biases and the $M$ case-specific valence probabilities are estimated from the data.

Fitting the AHPI model to the $N=63,115$ interactions yields a score $S_k$ for each of the $K=2,064$ law firms $k \in \{1, \ldots, K\}$, reflecting their underlying litigation performance (see Figure~\ref{fig:sketch} for an overview).  
To contextualize this ranking, we compare it against three widely used public rankings:  
\textit{Vault Law 100, ALM's Global 200}, and \textit{Embroker Top 300} from the year 2022.  
These rankings differ in methodology but do not incorporate litigation data, instead relying on factors such as reputation and firm size (see SI Appendix for details).  
To quantify the similarity between AHPI and official rankings, we compute the Kendall $\tau$ correlation coefficient (Spearman correlation yields similar results) for the subset of law firms appearing in both rankings.  
The correlations are effectively zero (Figure~\ref{fig:rank_clock}, top), suggesting that existing rankings may not reliably reflect empirical performance.  
The bottom plot of Figure~\ref{fig:rank_clock} compares AHPI to three public rankings for the 20 firms appearing in all rankings, visually underscoring their differences.

As a byproduct of the fitted law firm scores, we also estimate case biases $\{ \epsilon_m \}$ and valence probabilities $\{q_m\}$ (Table \ref{tab:case_summary}).
The case biases are consitently positive, in line with the observation that the defendant win rate is a priori higher. 
The valence probability $q_m$ quantifies the extent to which law firm rankings influence case outcomes for a case of type $m$.
A valence probability of $0.5$ suggests that case outcomes are effectively random with respect to law firm skill, whereas a value close to 1 indicates that law firm rankings strongly predict success (see Methods and Figure \ref{fig:sketch}).
Our results consistently yield high valence probabilities, reinforcing both qualitative and quantitative evidence on the critical role of legal representation~\cite{Szmer2007,Miller2015}.

\subsection*{Score-Based Prediction of Trial Outcomes}

A critical aspect of legal strategy is the assessment of the likelihood of litigation success, a challenge that has been extensively studied both theoretically and empirically~\cite{Priest1984,Shavell1996,Chang2021}.  
With the rise of big data and machine learning, predictive models have become increasingly quantitative, allowing data-driven evaluations of case viability~\cite{Ruger2004,Zhong2018,Medvedeva2020}.  
These methods have gained particular traction in the litigation finance industry, where portfolios of lawsuits are systematically analyzed for their potential returns~\cite{Molot2010,Lera2022}.  

Here, we examine whether our law firm rankings can enhance the prediction of litigation outcomes.
Recall from above that we use the first 80\% of all cases to fit the $K=2,064$ law firm scores $\{ S_k \}$, as well as $M=5$ case biases $\{\epsilon_m\}$ and valence probabilities $\{ q_m \}$.
We now use the remaining 20\% to predict the outcome of lawsuits.

\begin{figure}[!htb]
    \centering
    \includegraphics[width=\linewidth]{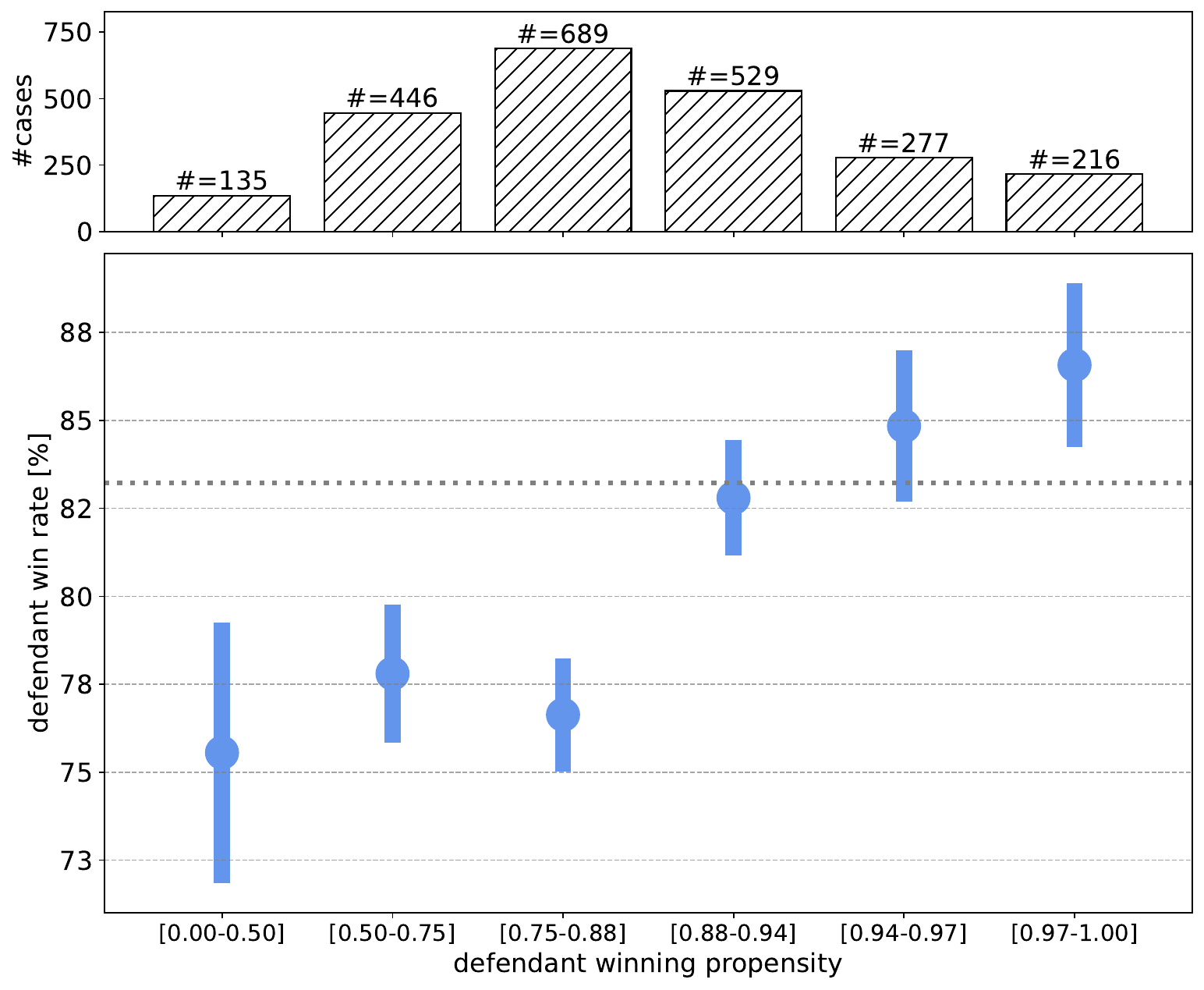}
    \caption{
            Predicted propensity of a defendant win for out-of-sample test cases, grouped into six bins.
            \textbf{(Top)} Number of cases within each bin.  
            \textbf{(Bottom)} 
            Winning propensity based on the AHPI ranking compared to actual defendant win rate in each bin.
            Error bars indicate standard deviations computed via 100 bootstrap resamples.
            The dotted line corresponds to the 83\% baseline defendant win-rate across all cases.
            Empirical defendant win-rates for cases with low propensities are significantly below the baseline while win-rates for cases with high propensities are significantly above the baseline, highlighting the model's ability to predict case outcomes based on law firm rankings.  
            }
    \label{fig:predictions}
\end{figure}

In principle, we could predict that law firm $A$ wins against law firm $B$ if $S_A > (S_B+\epsilon_m)$, and vice versa.
But because defendants win most of the cases, the naive baseline strategy that always just predicts that the defendant wins achieves a difficult-to-beat overall accuracy of $83\%$. 
To show that our ranking provides additional information beyond such a naive guess, we take advantage of the probabilistic nature of our score-based model.  
Rather than making binary win/loss predictions based solely on whether $S_A > S_B + \epsilon_m$, we account for the fact that the propensity of a plaintiff to win is a continuous function of the score difference. 
The fitted valence probabilities $q$ are consistently close to 1 so the sigmoid of the bias-adjusted score difference can be directly interpreted as a winning propensity (see Methods and Figure \ref{fig:sketch}).  
Therefore, the larger the difference $S_A - (S_B+\epsilon_m)$, the higher the predicted propensity that law firm A prevails.  
In contrast, when $S_A$ and $S_B+\epsilon_m$ are similar, the predicted propensity of either party to win approaches 50\%, reflecting a greater uncertainty in the case outcome.  
We therefore group our 20\% test cases into 6 bins with similar winning propensities and analyze the defendant win rate per bin. 
As shown in Figure~\ref{fig:predictions} the empirical defendant win-rates for cases with low propensities are significantly below the $83\%$ baseline while win-rates for cases with high propensities are significantly above the baseline.
This result demonstrates that our method
can predict case outcomes based on law firm rankings, enabling a far more granular prediction framework and allowing litigants to make more informed decisions in law firm selection and litigation strategy.

We can further compare the performance of our scores with those from the official rankings, where we use the published rankings as their score. 
Given the limited overlap among cases, we assess each ranking's performance on its specific batch of cases. 
To enable comparison between rankings, we reduce the frequency of the majority class (defendant wins), ensuring the data is balanced, meaning the defendant and plaintiff each win 50\% of the time. 
Consequently, a simple prediction yields an average accuracy of 50\%.
We then predict that law firm $A$ wins against law firm $B$ if $S_A > (S_B+\epsilon_m)$, and vice versa.
As illustrated in Figure~\ref{fig:rank_clock} (top), our approach surpasses this benchmark by nearly 10\%. 
In contrast, the leading result from the Embroker ranking shows an outperformance of less than 3\%, while Vault and AML fall short of even a random guess. 
Further, given that our ranking includes substantially more data (encompassing over 2,000 law firms), our performance displays smaller standard errors, derived through bootstrapping.
Taken together, these results suggest that a data-driven ranking provides superior predictive power for litigation outcomes.

\section*{Discussion}

\subsection*{Data-Driven Law Firm Rankings}

A growing body of empirical legal research shows that systematic analysis of court records can enhance legal practice and improve the administration of justice.
However, much of the legal community remains resistant to quantitative approaches, with skeptics only accepting data-driven insights if they are transparent, apolitical, and incontrovertible~\cite{Pah2020}.
In line with this perspective, we contribute to this effort by developing an objective ranking measure based on observed litigation outcomes rather than subjective prestige metrics.
By providing a data-driven evaluation of law firm performance, our approach aims to increase transparency and offer litigants a more reliable basis for selecting legal representation.

The application of our ranking approach offers practical benefits, particularly in enhancing the decision-making capabilities of litigants. 
It provides a quantitative basis for selecting law firms and devising litigation strategies, supporting more informed decisions such as whether to settle a case based on the litigation strength of the opposing law firm. 
Although well-resourced litigants may already possess a qualitative understanding of law firm performance, our approach democratizes access to these insights, making them available more broadly and equitably.
This shift could transform the way parties engage with legal representation and influence the dynamics of broader legal strategy.
We find that existing rankings, which are based on reputation and other extraneous factors, do not capture litigation success.
Our finding parallels similar results for physician rankings, which were shown to have little correlation with the \textit{US News \& World Report} medical school ranking~\cite{Tsugawa2018}.

\subsection*{Limitations \& Future Work}
\label{sec:limitations}

While our approach provides a novel computational framework for ranking law firms based on litigation outcomes, it is subject to certain limitations that warrant further discussion.  
These limitations primarily concern selection bias in our dataset and constraints imposed by available public data.

One of the primary concerns is survivorship bias, as our dataset includes only cases that resulted in a judicial opinion, excluding those that settled privately.  
Since most legal disputes are resolved through settlement rather than trial~\cite{prescott2016comprehensive}, our data represents only a subset of all legal interactions---cases where litigants were unable to reach an agreement before judgment.  
This limitation is mitigated by the fact that litigants negotiate `in the shadow of the law'~\cite{Cooter1982}, meaning that trial outcomes influence settlement behavior. 
If law firm strength significantly affects case outcomes, as our analysis suggests, firms with stronger litigation records may also achieve better settlement outcomes.  
While direct data on settlements remains scarce, future work should aim to incorporate settlement outcomes to construct a more holistic measure of law firm performance.  

Our ranking framework can be extended to account for settlements in three ways.  
First, settlements can be introduced as an explicit third outcome in our ranking model.
This approach aligns with extensions of the Bradley-Terry model that incorporate draws~\cite{agresti1990}, allowing us to infer law firm effectiveness in both litigated and negotiated outcomes.  
Second, if textual records of settlements become available, machine learning methods similar to those applied in this study could extract structured data, such as the plaintiff’s initial demand and the final settlement amount.  
Modeling settlement outcomes alongside trial results would refine our assessment of law firm performance beyond litigation alone.  
Third, a sufficiently large settlement dataset would enable analysis of the relationship between settlement likelihood and trial win probabilities,  
helping to estimate the degree to which law firm reputation and skill influence bargaining dynamics.  
These extensions would enhance the robustness of our ranking system and provide a more comprehensive picture of legal representation effectiveness.  

Selection biases represent another concern if litigants select the `best' law firms for the hardest cases. 
While it is difficult to gather empirical data to quantify such effects, we believe that they are not common in practice. 
Only highly sophisticated litigants, especially large corporations with legal departments, would have sufficient information about law firm performance and case merit to make such decision. 
In practice, companies have established relationships with a set of law firms~\cite{coates2011hiring} and it appears uncommon for them to hire a new firm to handle a particularly difficult case. 

In cases involving multiple plaintiff or defendant law firms, we reduced interactions to pairwise binary comparisons between individual firms.  
Future work could explore ranking methodologies that account for law firms working as teams,  
incorporating cooperative dynamics and the potential synergies between firms representing the same party. 
Similarly, we do not explicitly account for law firm mergers, acquisitions, or splits, which may affect the continuity of firm performance data.
Future work could incorporate historical firm linkages to adjust rankings dynamically in response to structural changes.  

Another limitation is that our analysis focuses on U.S. federal civil litigation, meaning our findings may not generalize to state courts or international jurisdictions.  
Federal district courts provide a structured and relatively consistent dataset, making them an ideal starting point, but differences in legal frameworks, procedural rules, and judicial discretion may yield different patterns elsewhere.  
Expanding our methodology to state courts and non-U.S. legal systems would provide valuable comparative insights and test the broader applicability of our ranking framework.  

If our ranking system gains widespread adoption, law firms may adjust their litigation strategies to improve their standing.
While a focus on a single metric can sometimes cause it to lose its effectiveness as a meaningful measure, 
our framework directly reflects litigation outcomes rather than proxies like reputation or revenue and thus naturally aligns with the interests of clients seeking effective legal representation.
Moreover, the nature of the AHPI ranking makes it difficult to manipulate as selecting only weak opponents would put a ceiling on a firm's ranking.
Of course, litigation success is only one dimension of legal practice.
Other factors, such as value-for-money, client satisfaction, and strategic case selection, are equally important.
Future iterations of our model could incorporate financial data, case duration, or fee structures to evaluate firms based on overall value rather than just win rates.
Additionally, refining rankings by incorporating judicial characteristics, such as judge-specific biases~\cite{Mahari2024},
could provide more granular insights into litigation dynamics and further improve decision-making for litigants.

Finally, while our model is designed for legal disputes, its underlying framework can be applied to other adversarial settings. 
The asymmetric, heterogeneous pairwise interactions (AHPI) algorithm is broadly applicable to competitive environments where entities engage in repeated interactions, such as political races, lobbying efforts, or corporate competition.  
Extending our ranking system to multi-party legal disputes or auction-based legal markets could further enhance its applicability and provide deeper insights into legal strategy and competition.  

Despite these limitations, our work establishes an empirical foundation for evaluating law firm performance based on objective litigation outcomes.  
By shifting the focus from reputation-based rankings to data-driven assessments, we provide a framework that enhances transparency in the legal profession and lays the groundwork for future research in empirical legal studies. 

\subsection*{Conclusions}

We introduce a data-driven ranking of law firms based on litigation outcomes, outperforming widely used reputation-based and profit-oriented rankings in predicting case results.  
By prioritizing performance over prestige, our ranking provides a more reliable tool for selecting legal representation, helping litigants make informed decisions, and leveling the playing field for less experienced parties.  
To construct this ranking, we compile a large dataset of U.S. federal trial records from open-source judicial opinions, enabling systematic analysis of law firm performance at scale.  
These findings establish a foundation for more transparent, data-driven assessments of legal performance.  
As access to legal data improves, future extensions of our framework, incorporating settlements, jurisdictional variations, and financial metrics, could fundamentally transform the way law firms are evaluated, reshaping both legal practice and access to justice.  

\section*{Data Availability}

The source data for our work are based on \cite{Mahari2024} and augmented by identifying and disambiguating law firms and their roles. 
The data for reproducing all the results will be made available upon publication. 
The analyses and rankings that we report may be reproduced by running our code.

\section*{Code Availability}

The code to reproduce all the analyses in this paper can be found on GitHub at \url{https://github.com/mojona/law_firm_ranking}.

{
\balance
\tiny
\bibliographystyle{elsarticle-num}
\bibliography{bibliography.bib}
}

\section*{Competing interests}

The authors declare no competing interests.

\section*{Methods}
\label{sec:Methods}

\subsection*{Data}
\label{sec:data}

Our dataset is constructed from judicial opinions provided by the Case Law Access Project (CAP), which contains textual records of U.S. Federal District Court cases, closely following the method presented by ref.~\cite{Mahari2024}.
We restrict our analysis to civil litigation in federal district courts, excluding appellate and Supreme Court cases to ensure a consistent trial-level dataset.
This leaves us with 302,988 cases in the form of judges' published opinions. 
These opinions include detailed case descriptions but lack structured metadata on case type, case outcome, or legal representation.  
To address this, we leverage structured data from the Integrated Database (IDB) provided by the Federal Judicial Center (FJC) as training data to develop machine learning models that infer key case attributes directly from judicial opinions. 

The IDB is a structured database of federal civil lawsuits, containing metadata fields such as case filing date, case type, and case outcome (plaintiff win, defendant win, or unknown).  
By matching CAP cases to IDB records via docket numbers and filing dates, we successfully align 31,222 cases between the two datasets~\cite{Mahari2024}.
These matched cases serve as labeled training data for extracting structured information from the full CAP corpus, which otherwise lacks metadata annotations.  

To infer case outcomes and types for the CAP cases, we fine-tune two BERT-based classifiers using the 31,222 IDB-mapped cases as training data.
Both classifiers take as input the judge's opinion.
The first model is a binary classifier that predicts whether the plaintiff won or lost, 
while the second is a five-class classifier that assigns cases to one of five major case types---civil rights, contracts, labor, torts, and \textit{other} (Table~\ref{tab:case_summary}).  
Once trained, these models are applied to the filtered CAP dataset, allowing us to classify case outcomes and case types for all 302,988 cases.
After discarding cases for which our predictive models have low confidence, we obtain a total of 205,454 labeled cases.
This dataset forms the basis of our analysis. 
We refer to the SI Appendix for an in-depth discussion of how these models have been trained. 

While this procedure provides structured labels for case type and outcome, it does not yet identify the law firms involved in each case.  
In CAP, law firm names appear in the attorney sections of judicial opinions, often inconsistently formatted due to OCR errors, abbreviations, or typographical variations (e.g., `Putman \& Putman' vs. `Put-Man and Put-Man').  
To extract this information, we apply a pattern-matching approach to detect law firm names, followed by hierarchical clustering based on Levenshtein distance to merge name variants.  
A similar method is used to determine which law firms represent plaintiffs and defendants in each case.  
This step ensures consistency in firm identification and representation assignment and is detailed further in the SI Appendix.
After discarding cases in which we cannot identify at least one law firm on each side, we are left with a total of 67,619 cases and 64,656 law firms.
A detailed breakdown of these numbers is found in the SI Appendix. 

A key feature of our dataset is that it enables the construction of an \textit{interaction network} of law firms. 
Each law firm is represented as a node and each lawsuit interaction is represented as an edge. 
This network exhibits considerable variation in density, as some law firms engage in frequent litigation while others appear only sporadically.  
To systematically analyze this structure, we introduce the \textit{$Q$-factor}, which quantifies the average number of observed cases per law firm within a given subset of the network.  
By definition, the $Q$-factor of our total dataset is $167,439/54,541 \approx 3.1$. 
By iteratively removing the law firms with least interactions, we trim the network so that the average number of interactions per law firm, i.e. the $Q$-factor, is increasing. 
Setting $Q$ sufficiently high ensures that each firm has a meaningful number of observed interactions, allowing for a more reliable estimation of its litigation performance.  
For the results presented in Figures \ref{fig:rank_clock} and \ref{fig:predictions}, we use $Q=30$ to compute law firm rankings, 
yielding an interaction network with $N=63,115$ interactions involving $K=2,064$ law firms. 
We confirm in the SI Appendix that our results remain robust across a range of $Q$ values from $20$ to $55$. 

\subsection*{Modeling Asymmetric, Heterogeneous, Pairwise Interactions}
\label{sec:asym_heter_inter}

The ranking of entities based on pairwise interactions is an ubiquitous problem that arises in domains such as
information retrieval~\cite{Page1999}, 
decision theory~\cite{SAATY2008}, 
sports~\cite{glickman1999}, 
and university rankings~\cite{Avery2013}.  

Several algorithms have been established to estimate rankings from observed outcomes.  
The Elo rating system, widely used in chess~\cite{elo1960}, updates player scores dynamically based on match outcomes, 
where score adjustments depend only on the pre-match ratings of the two competitors.  
Although Elo provides an efficient heuristic for continuously updated rankings,  
it does not maximize a formal likelihood function and cannot easily incorporate multiple interaction types or systematic biases in matchups.  
In contrast, AHPI represents the probability that one competitor defeats another as a logistic function of the difference in their latent skill levels~\cite{BradleyTerry1952},  allowing for a principled estimation of rankings through likelihood maximization.  
Recent extensions of this model have incorporated multiple interaction types while also enhancing computational efficiency~\cite{Newman2022}.  
We propose the \textit{asymmetric heterogeneous pairwise interactions} (AHPI) ranking algorithm,  
which further generalizes this approach to account for asymmetric games, where one entity has a higher a priori chance of winning (e.g., a `home field advantage' for defendants in litigation).  
This formulation allows for a more statistically rigorous ranking framework, ensuring that rankings reflect both observed outcomes and structural biases in competition.

We consider $K$ entities (e.g. law firms) $k=1,...,K$ competing in a total of $N$ pairwise interactions (e.g. lawsuits) $n=1,...,N$.
Each interaction is of one of $M$ types (e.g. lawsuit case types, or variations of the game) $m=1,...,M$.
The interaction is asymmetric in the sense that defendants are more likely to win a case.
We assume that this `home field advantage' or `defendant bias' is specific to the case type and denote it by $\epsilon_m$.
The AHPI algorithm assigns a score $S_k$ to every competing entity $k$.
Consider the interaction $n \in \{1, \ldots, N \}$ that is of type $m$ and involves the entity $A$ and the privileged entity $B$ with respective scores $S_A$ and $S_B$.
The outcome is modeled in two stages:
First, a favored entity is determined. 
The probability that $A$ is favored is given by 
    \begin{subequations} \label{eq:winning_probas}
    \begin{equation}
        \rho_n(A) = \frac{1}{1+\exp(-(S_A-(S_B+\epsilon_m))}
        \label{eq:favoured_probability}
    \end{equation}
where the privilege $\epsilon_m$ skews this probability in $B$'s favor. 
It is implicitly understood that the interaction type $m$ is dependent on the game $n$, that is, $m=m(n)$. 
We denote $\epsilon_m$ for brevity and proceed similarly for all subsequent quantities. 
Second, the winner of the interaction is determined by introducing the valence probability $q_m$.
The valance probability captures the probability that the favored entity wins. 
Hence, $A$ wins with probability
    \begin{equation}
        p_n(A) = q_m \cdot \rho_n(A) + (1-q_m) \cdot \left(1-\rho_n(A)\right)
        \label{eq:winner_probability}
    \end{equation}
    \end{subequations}
and $B$ wins with probability $1 - p_n(A)$.
The advantage of this two-step approach is that the skills of the entities, expressed in their scores $S$, are decoupled from the probabilistic outcome associated with interaction type $m$ through $q_m$.
If the score difference corrected by the privilege in \eqref{eq:favoured_probability} is 0, both entities have equal probability of being favored.
To what extent being favored implies winning is subsequently determined by the valence probability $q_m$.
A value of $q_m$ close to 1 indicates that the favored entity is highly likely to win, whereas $q_m=0.5$ indicates that being favored does not have an impact.
This approach has previously been used to calibrate the relative contributions of skill and luck \cite{Jerdee2023}.
We find that fitted valence probabilities tend to be close to $1$, suggesting that the law firms' score differences are highly indicative of case outcomes. 

\subsection*{Estimation of Latent Scores}
\label{sec:estimate}

We now turn to the estimation of the scores $\{ S_k \}_{k=1}^K$, privileges $\{ \epsilon_m \}_{m=1}^M$ and valence probabilities $\{ q_m \}_{m=1}^M$ given $N$ observed interactions $\{ I_n \}_{n=1}^N$.
To this end, we introduce for every interaction $n$ the stance variable $\sigma_n$ which takes a value of 1 if the favored entity is the winner and 0 else. 
We label the winning entity by $u_n$ and the losing one by $v_n$. 
It is convenient to define $\lambda_{k} \equiv e^{S_{k}}$.
To further simplify notation, we use $S$ instead of $\{S_k\}_{k=1}^K$ to denote the set of scores, and similarly for $\{ \sigma_n \}, \{ \epsilon_m \}$ and $\{ q_m \}$.
In order to tackle potential convergence issues, we use a Bayesian framework and introduce logistic priors for scores and privileges \cite{Whelan2017,Newman2022}.
Applying Bayes' theorem, the likelihood 
$P( s, q, \sigma, \epsilon \mid x )$
is
\begin{equation} \label{eq:likelihood}
\prod_{n} P_{n}(x,\sigma \mid \lambda_{u},\lambda_{v},\epsilon,q) \prod_{k=1}^{K}\frac{\lambda_{k}}{(\lambda_{k}+1)^{2}} \prod_{t=1}^{M}\frac{1}{(e^{\epsilon_m}+1) (e^{-\epsilon_m}+1)}
\end{equation}
where the second and third product terms stem from the prior on the scores and privileges, respectively. 
By defining the privilege stance variable $c_n$ as taking value -1 if the winner was privileged and 1 else, 
and by using the definitions of the probabilities $\rho_n$ and $p_n$ in \eqref{eq:winning_probas}, 
it holds that
\begin{equation} \label{eq:P_r_conditional}
P_n(x,\sigma \mid \lambda_{u},\lambda_{v},\epsilon,q)=\frac{(e^{c_n\cdot\epsilon_m}\cdot\lambda_{u}\cdot q_m)^{\sigma}\cdot(\lambda_{v}(1-q_m))^{1-\sigma}}{e^{c_n\cdot\epsilon_m}\lambda_{u} + \lambda_{v}}.
\end{equation}
It turns out that directly maximizing (the logarithm of) the likelihood \eqref{eq:likelihood} is challenging, and an expectation maximization algorithm is better suited \cite{Newman2022}.
To this end, we first note that across $N$ games, there is a total of $R = 2^N$ outcomes of the binary stance variables $\{ \sigma_n \}$, and we denote by $\Pi$ any probability distribution over these outcomes.
By Jensen's inequality, it holds that
\begin{equation} \label{eq:Jensen}
    \log\sum_{r=1}^{R}P(S,q,\sigma,\epsilon \mid x)
    \geqslant 
    \sum_{n=1}^{N}\Pi(\sigma)\log\frac{P(S,q,\sigma,\epsilon \mid x)}{\Pi(\sigma)}.
\end{equation}
If the right-hand side equals the left-hand side, we can work with the sum of logarithms which renders the calculations below analytically tractable. 
For fixed $S$, $\epsilon$ and $q$, equality holds for a specific probability distribution $\Pi$: 
\begin{subequations} \label{eq:Pi_specific}
\begin{equation}\label{equ:full_pi}
\Pi(\sigma) =\prod_{n=1}^{N}\pi_n^{ \sigma_n } ~(1-\pi_n)^{1-\sigma_n}
\end{equation}
where
\begin{equation}\label{equ:final_pi_r}
\pi_n 
= 
\frac{e^{c_n \cdot \epsilon_m} \cdot \lambda_{u_n} \cdot q_m}{ \lambda_{u_n} \cdot e^{c_n \cdot \epsilon_m} \cdot q_m+\lambda_{v_n \cdot (1-q_m)} }
\end{equation}
\end{subequations}
can be interpreted as the posterior probability that $u_n$ is the favored entity \cite{Newman2022}.

We now apply expectation maximization to the right hand side of \eqref{eq:Jensen} with $\Pi$ given by \eqref{eq:Pi_specific}.
To maximize expectations, we set the derivative of the right-hand-side of \eqref{eq:Jensen} with respect to $S, \epsilon$ and $q$ equal to zero while holding the distribution over $\sigma$ constant. 
This results in the following set of equations where $\delta_{\mu,\nu}$ is the Kronecker delta.
\begin{subequations}
\begin{align}
    q_m         &=  \frac{\sum_{n=1}^{N}\delta_{m_n,m} \cdot \pi_n}{\sum_{n=1}^{N}\delta_{m_n,m}} \label{eq:max_for_q} \\
    0           &=  \frac{1-e^{\epsilon_m}}{1+e^{\epsilon_m}} + \sum_{n=1}^{N} \delta_{m_n,m} \cdot \left[ \pi_n c_n-\frac{\lambda_{u_n} \cdot e^{c_n \epsilon_m}}{\lambda_{u_n} \cdot e^{c_n \epsilon_m}+\lambda_{v_n} }c_n \right] \label{eq:max_for_eps} \\
    \lambda_k   &= \left[1+\sum_{n=1}^{N}\delta_{u_n,k} \pi_{n}+ \delta_{v_n,k} (1-\pi_n)\right] \cdot \notag \\ 
                &~~~~\left[ 
                        \frac{2}{1+\lambda_k} 
                        +
                        \sum_{n=1}^{N}
                        \frac{ 
                        \delta_{u_n,k} \gamma_n ( \gamma_n \lambda_{u_n} +\lambda_k ) + \delta_{v_n,k} (\gamma_n \lambda_k + \lambda_{v_n})
                        }                        
                        {
                        (\gamma_n \lambda_k + \lambda_{v_n}) ( \gamma_n \lambda_{u_n}+\lambda_k)
                        }
                        \right]^{-1} 
                        \label{eq:max_for_lam}
\end{align}
\end{subequations}
Equation \eqref{eq:max_for_q} yields an explicit expression for the valence probability $q_m$ where $\pi_n$ is given by \eqref{equ:final_pi_r}.
Equation \eqref{eq:max_for_eps} yields an implicit expression for the privilege $\epsilon_m$, which can be solved numerically. 
Equation \eqref{eq:max_for_lam} yields a nested set of equations for the exponential scores $\lambda_k$, which can be solved iteratively until convergence is reached. 

Finally, it should be noted that the above calculations are invariant under the mapping
\begin{equation}\label{eq:symmetry}
    \left\{
    \begin{aligned}
        &f: \mathbb{R} \times \mathbb{R} \times [0,1] \rightarrow \mathbb{R} \times \mathbb{R} \times [0,1] \\
        &f(S, \epsilon, q) = (-S, -\epsilon, 1-q)
    \end{aligned}
    \right.
\end{equation}
which suggests that the ranking might be `inverted'.
This symmetry can be broken by analyzing the values of the valence probabilities.
If for an interaction type $m$ a higher rank is assumed to imply a higher winning probability, but if $q_m<0.5$, then the ranking must be inverted according to \eqref{eq:symmetry}.
Likewise, if a higher rank is assumed to imply a lower winning probability and $q_m>0.5$, then the ranking must be inverted.

\subsection*{Fitting Latent Scores on Dense Data Subsets}
\label{sec:fit}

Our dataset contains 167,439 interactions (\textit{games}) across 54,541 law firms (\textit{entities}).
Each interaction is of one of five types (Table \ref{tab:case_summary}).
Ideally, we assign a ranking score $S_k$ to each law firm $k \in 1, \ldots, K=54,541$ by applying the above expectation maximization algorithm. 
However, it is intuitively plausible that the fewer interactions are observed per entity, the less reliable the estimated scores (see SI Appendix for a demonstration on data with known ground truth).
We therefore systematically trim our dataset to a subset of sufficient interactions via $Q$-factor, described above.  
By definition, the $Q$-factor of our total dataset is $167,439/54,541 \approx 3.1$.
In order to find a subset of the data with sufficiently high target $Q$-factor, 
we iteratively remove the entities with lowest number of interactions and recalculate the $Q$-factor.
We stop once the average number of interactions in the remaining data is larger or equal to $Q$. 
Law firm rankings (Figure \ref{fig:rank_clock}) and outcome predictions (Figure \ref{fig:predictions}) have been calculated on a subset with $Q=30$ for which the number of law firms is reduced to 2,064 and the number of interactions is reduced to 63,115.
Qualitatively similar results for a larger range of $Q$-factors are found in the SI Appendix. 

For a given dataset of interactions with fixed $Q$-factor, we proceed with the estimation of the model parameters. 
Concretely, we estimate the law firm scores $S_k = \log(\lambda_k)$, the case type biases $\epsilon_m$ and the valence probabilities $q_m$ via the above described expectation maximization algorithm. 
The following initial values are used for all firms and interaction types, respectively:  $\lambda_k=0.9$, $q_m=0.5$, $\epsilon_m=0$.
The algorithm is iterated until convergence is reached, that is once the correlation between the ranking of two subsequent iterations is above 99.9\% and the maximum absolute change in any ranking score, valence probability and case type asymmetry is below 0.01.

The parameters are only fitted on the first 80\% of the cases according to their temporal order (accordingly, the interaction network and the $Q$-factor were determined only on the training data, not the test data).
The remaining 20\% are used to predict defendant winning probabilities via \eqref{eq:winning_probas} where $S_k, \epsilon_m$ and $q_m$ are replaced by their fitted best estimates $\hat{S}_k, \hat{\epsilon}_m$ and $\hat{q}_m$, respectively. 
In accordance with the defendant bias (Table \ref{tab:case_summary}), we estimate consistently positive $\hat{\epsilon}_m$ parameters equal to $1.90, 1.66, 0.32, 2.00$ and $2.03$ for `other', `contract', `torts', `labor' and `civil rights' cases, respectively. 
The valence probabilities are found to be very close to or even equal to $1.0$, implying that the favored entity typically wins.
Similarly high values have been observed in previous work \cite{Newman2022} and we refer to the SI Appendix for additional discussions. 
Qualitatively, this suggests that the outcome of a lawsuit is strongly influenced by the difference of skill-levels of the competing law firms. 

\newpage
\onecolumn
\appendix
\begin{center}
    \LARGE \bfseries SI Appendix
\end{center}
\renewcommand{\thefigure}{\arabic{figure}} 
\renewcommand{\thetable}{\arabic{table}}  
\renewcommand{\figurename}{Supplementary Figure} 
\renewcommand{\tablename}{Supplementary Table}   
\setcounter{figure}{0} 
\setcounter{table}{0}  

\section{Transformer-Based Case Labeling}
\label{SI:transformer_labeling}

As described in the Data section of the main article, our dataset is built upon the Harvard Case Law Access Project (CAP),  
which provides judicial opinions detailing case decisions.  
To systematically extract structured information from these texts, we follow the methodology described in Ref.~\cite{Mahari2024} to identify two key attributes:  
first, whether the plaintiff won or lost the case, and second, the case type classification.  
To accomplish this, we fine-tune two Longformer models~\cite{Beltagy2020},  
one dedicated to case outcome classification and the other to case type identification.  
Below, we outline the details of how these transformer-based models were incorporated into our dataset.

While natural language processing presents technical challenges, this particular task is relatively straightforward from an information-theoretic perspective, as both the case type and outcome are explicitly mentioned within judicial opinions.  
This stands in contrast to our primary contribution,  
which involves predicting case outcomes based on law firm rankings rather than directly extracting factual case details from judicial texts after the case has concluded.

\subsection{Case Outcome Classification}

\begin{figure*}[htb]
    \centering
    \includegraphics[width=0.8\linewidth]{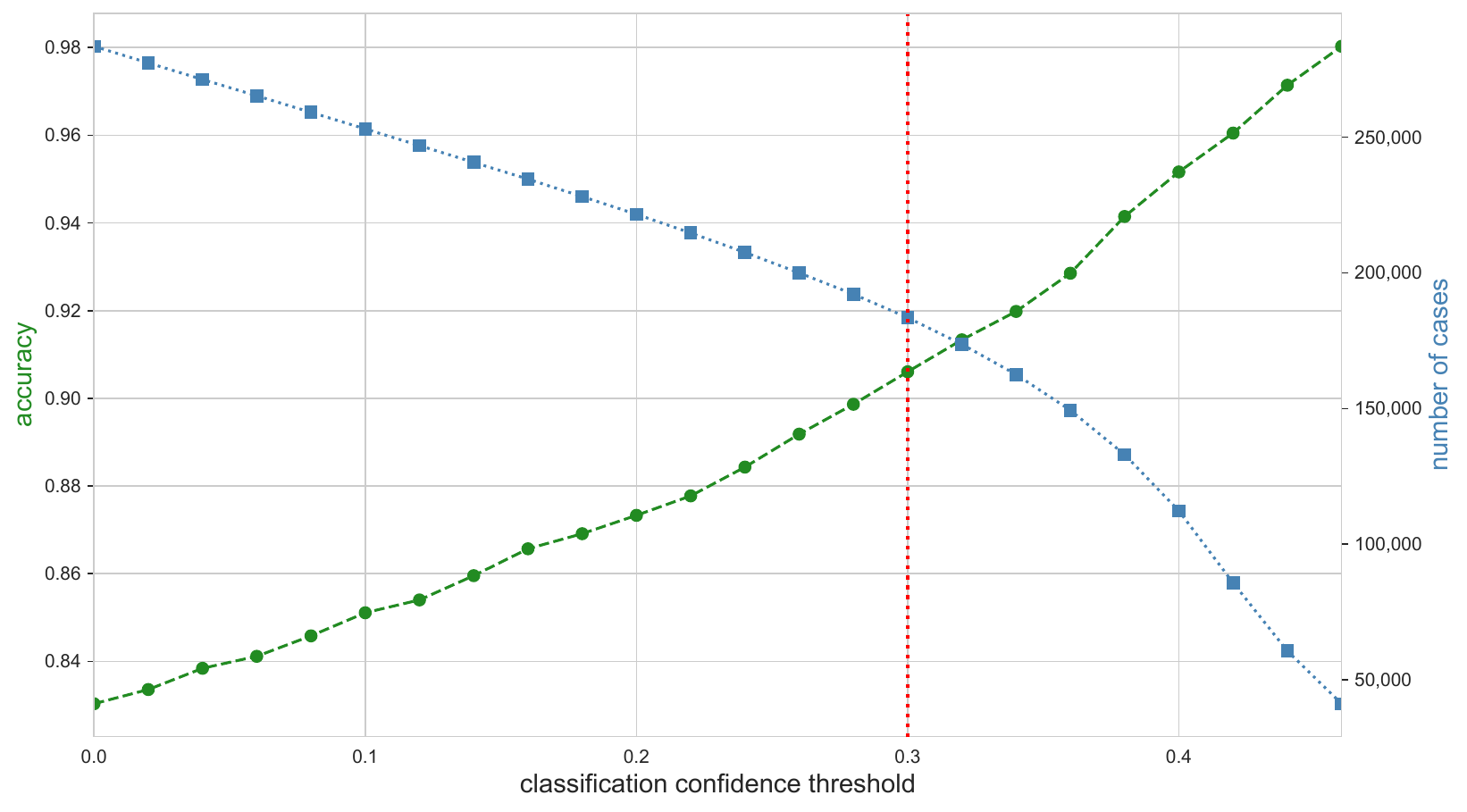}
    \caption{
    Trade-off between data augmentation accuracy and dataset size.
    Accuracy on test data (left y-axis) and number of retained cases (right y-axis)  
    as a function of the classifier confidence threshold $\tau$.
    }
    \label{fig:transformer_classifier_sensitivity}
\end{figure*}

We fine-tuned a transformer-based classifier to map a judge’s opinion to a binary outcome:  
whether the plaintiff won (1) or lost (0).  
For training, we use 31,222 cases matched to the IDB dataset, reserving 10\% of this data for validation.
Rather than producing a strict binary classification,  the trained model generates a sigmoid output $p \in [0,1]$,  
which represents the classifier’s confidence in predicting a plaintiff victory.  
A naive approach would classify a case as a plaintiff win if $p > 0.5$ and a loss otherwise.  
However, this threshold does not account for the uncertainty near $p \approx 0.5$,  
where misclassification rates are likely to be higher.  
To improve classification reliability,  
we introduce a confidence threshold $\tau \in (0,0.5)$ and only retain cases where $| p - 0.5 | > \tau$.  
This filtering approach ensures that we consider only cases where the classifier has a sufficiently high confidence level.  
Supplementary Figure~\ref{fig:transformer_classifier_sensitivity} illustrates the trade-off between accuracy and dataset size as a function of $\tau$.  
While accuracy is already high for small values of $\tau$,  
we select $\tau = 0.3$ to balance data quality and sample size.  
At this threshold, the classifier achieves an accuracy of over 91\% on the validation set, 
leaving us with a total of 205,454 confidently labeled cases.
Thus, we only retain cases where the model predicts an outcome with either $p < 0.2$ or $p > 0.8$ and discarding all others. 

\subsection{Case Type Classification}

A similar procedure is employed for the case type classifier, which predicts the case’s legal category using a softmax distribution over five case types.  
The model assigns probability scores to each possible type:  
\( p_{\text{civil rights}}, p_{\text{contract}}, \dots, p_{\text{torts}} \),  
where probabilities sum to one. 
If all case types were assigned equal likelihoods, each would receive a probability of approximately $20\%$.
However, we observe that for each prediction, there is consistently a case type with a predicted probability above 50\%. 
We thus assign to each CAP opinion the case type with the hightest predicted probability.

\subsection{Summary}
\label{SI:transformer-summary}

We begin with an initial set of 302,986 judicial opinions from the CAP dataset.  
Among these, 31,222 cases are matched with the IDB dataset, providing ground-truth labels for case type and outcome.  
Using this labeled subset, we fine-tune two transformer-based models:  
one to classify case types and another to determine whether the plaintiff won or lost.  
Once trained, these models are applied to the full CAP dataset, excluding the training cases.  
To ensure high data quality, we discard cases where the classifier's confidence falls below a predefined threshold.  
After this filtering step and adding back the training data, 
we retain a final dataset comprising 205,454 cases.
We refer to Supplementary Table \ref{tab:data_summary} for an overview.

\begin{table}[htbp]
\centering
\begin{threeparttable}
\caption{Summary of the different number of civil cases, law firms, and law firm interactions.}
\label{tab:data_summary}
\begin{tabular}{l|rrr|rrr|rrr}
\hline
 & \multicolumn{3}{c|}{Number of Cases} & \multicolumn{3}{c|}{Interactions} & \multicolumn{3}{c}{Number of Law Firms} \\
\cline{2-10}
 & Train & Test & Total & Train & Test & Total & Train & Test & Total \\
\hline
CAP & - & - & 302,988 & - & - & - & - & - & - \\
CAP with confidence & - & - & 205,454 & - & - & - & - & - & - \\
CAP with confidence and firms & - & - & 67,619 & - & - & - & - & - & 64,656 \\
\hline
$Q=3.1^*$ & 54,095 & 6,445  & 60,540  & 167,439 & 22,858 & 190,297 & 54,541 & 4,198 & 54,541 \\
$Q=5$  & 39,138 & 5,525  & 44,663  & 143,813 & 20,635 & 164,448 & 28,084 & 3,271 & 28,084 \\
$Q=10$ & 23,798 & 4,224  & 28,022  & 109,951 & 17,418 & 127,369 & 10,075 & 2,082 & 10,075 \\
$Q=15$ & 17,637 & 3,566  & 21,203  & 94,379  & 15,433 & 109,812 & 5,953  & 1,594 & 5,953 \\
$Q=20$ & 14,042 & 3,114  & 17,156  & 84,325  & 13,831 & 98,156  & 4,215  & 1,308 & 4,215 \\
$Q=25$ & 11,124 & 2,733  & 13,857  & 74,333  & 12,347 & 86,680  & 2,973  & 1,062 & 2,973 \\
$Q=30$ & 8,687  & 2,292  & 10,979  & 63,115  & 10,270 & 73,385  & 2,064  & 805   & 2,064 \\
$Q=35$ & 7,139  & 2,001  & 9,140   & 55,090  & 8,844  & 63,934  & 1,574  & 657   & 1,574 \\
$Q=40$ & 5,368  & 1,669  & 7,037   & 44,620  & 7,246  & 51,866  & 1,110  & 512   & 1,110 \\
$Q=45$ & 4,244  & 1,416  & 5,660   & 33,567  & 5,420  & 38,987  & 744    & 381   & 744 \\
$Q=50$ & 3,467  & 1,209  & 4,676   & 25,902  & 4,346  & 30,248  & 518    & 279   & 518 \\
$Q=55$ & 2,168  & 790    & 2,958   & 11,289  & 2,327  & 13,616  & 205    & 143   & 205 \\
\hline
\end{tabular}
\begin{tablenotes}
    \item[*] 
    We use an 80\%-20\% train-test split on the 67,619 cases, resulting in 54,095 cases for training and 13,523 for testing. 
    However, we only consider test cases for which we have a fitted score for both law firms, and so the number of test cases is reduced to 6,445. 
    This explains why the total number of cases is reduced from 67,619 to 60,540. 
    Similar remarks apply for the total number of law firms. 
\end{tablenotes}
\end{threeparttable}
\end{table}

\clearpage
\section{Extracting Law Firm Names and Roles}
\label{SI:bio_classification}

Beyond the judge’s textual case descriptions, the Case Law Access Project (CAP) also provides details on the law firms involved and their respective party representation.

\subsection{String Matching}

Each CAP case includes a list of attorney-related text entries, which we refer to as `attorney substrings.’
For instance, a case with list
\begin{center}
\texttt{[`Michael H. Auen, of Foley \& Lardner, Madison, Wis., for plaintiff.', `David C. Rice,  Patton Boggs, L.L.P., Madison, Wis., for defendants.']}.
\end{center}
consists of the two attorney substrings 
\begin{center}
\texttt{`Michael H. Auen, of Foley \& Lardner, Madison, Wis., for plaintiff.'}
\end{center}
and
\begin{center}
\texttt{`David C. Rice,  Patton Boggs, L.L.P., Madison, Wis., for defendants.'}
\end{center}
For this exemplified case, our extraction procedure should yield the following two tuples:
\begin{center}
\texttt{(`plaintiff', [`foley \& lardner']), (`defendant', [`patton boggs'])}. 
\end{center}

We now provide a brief overview of how law firm names and roles are extracted. 
More detailed information can be found in the Python code that is available at \url{https://github.com/mojona/law_firm_ranking}.

\begin{enumerate}

\item 
We apply basic string processing.
Frequent expressions containing full-points are replaced by their written-out equivalents (e.g. `U.S.' is replaced by `United States').

\item 
We discard cases with fewer than two attorney substrings since we require information for both the defendant and the plaintiff. 

\item 
In the vast majority of attorney substrings, the role of a law firm follows on the expression ` for ' (e.g. `for plaintiff').
Therefore, the expression following on ` for ' is read out.

\item \label{item:roles}
We are interested in cases where every law firm is mapped to a unique (defendant- or plaintiff-) role:
If more than one role has been extracted from an attorney substring, this case is discarded.

\item \label{item:firms_and}
Law firms containing the string ` \& ' are read out by extracting the term around ` \& ', for example from `... of Foley \& Lardner, Madison ...' we extract `Foley \& Lardner'.

\item \label{item:firms_abb}
Based on inspection of the CAP data and open source information of law firms, we compile a list of keywords such as `L.L.P.'.
Law firms containing such a keyword are read out by extracting terms around them.

\item 
The terms extracted for the law firms in steps \ref{item:firms_and} and \ref{item:firms_abb} are set to lowercase.
To account for anomalies and Optical Character Recognition (OCR) errors, we alter the terms by removing or replacing specific strings in specific positions.
For example, the string `$<$\&' is replaced by `\&', hereby presumably correcting an OCR error.

\end{enumerate}

To illustrate further, consider the District of Maine case from 1998 available at  
\url{https://case.law/caselaw/?reporter=f-supp&volume=998&case=0001-01}.  
The CAP record for this case includes the following two attorney substrings:  

\begin{itemize}
    \item `Ronald E. Colby, III, Sumner H. Lipman, Robert J. Stolt, Laura J. Garcia, \textbf{Lipman \& Katz}, Augusta, ME, \underline{for Plaintiff}.'
    \item `Richard G. Moon, Robert M. Hayes, \textbf{Moon, Moss, McGill \& Bachelder, P.A.}, Portland, ME, \underline{for Defendants}.'
\end{itemize}  

Using the above pattern matching approach, our extraction algorithm correctly identifies the law firms highlighted in bold and assigns them to their respective party roles.  
This results in a \textit{pairwise interaction} between `Lipman \& Katz' and `Moon, Moss, McGill \& Bachelder.' 
If multiple law firms were associated with a single party, each firm would be paired separately against firms on the opposing side.  
Extensions to account for multiparty interactions are discussed in the Limitations Section of the main article. 

Despite the robustness of our approach, the extraction of law firms is not always successful due to inconsistencies in the lawyer listings and OCR-related errors.  
Overall, we successfully extracted at least one law firm name from 65.3\% of all `attorney substrings.  
The remaining 34.7\% consist mainly of cases where the names of the law firms are missing or are formatted in a way that does not match common patterns.  

Although our dataset is necessarily constrained by the availability and formatting of legal records,  
our ranking approach is adaptable to more comprehensive datasets as they become available.  
As more data becomes accessible, the scope and accuracy of our empirical legal analysis can be further improved.
The computational framework developed here provides a foundation for such future enhancements.  

\subsection{Clustering Law Firm Names}

Above, we have described how we extract for each case a tuple of the form `(law firm name, role)'. 
However, due to OCR errors and spelling mistakes, we cannot directly associate each law firm name or role as its unique identity 
(e.g. `putman $\&$ putman' vs. `put-man $\&$ put-man').
We thus apply an agglomerative clustering mechanism which groups together similar strings. 
We separately apply this procedure for both the law firm names and the roles. 
Here, we focus on law firms, but the procedure for the roles is similar, albeit simpler. 
We refer to our Python implementation for more details. 

To assign strings to clusters, we must first define a metric that measures the distance between two strings.
We use the Levenshtein distance which counts how many transformations of a single character (insertions, deletions, or substitutions) have to be made to transform one string into another one.
For example, $\text{lev}(\text{plaintiff},\text{plaintif})=1$ and $\text{lev}(\text{plaintiff},\text{defendant})=8$.
More formally, the Levenshtein distance between two strings $a$ and $b$ is defined as: 
\[ \text{lev}(a,b) =
\left\{
\begin{array}{ll}
      \left| a \right| & \text{if $\left| b \right|$=0 }\\
      \left| b \right| & \text{if $\left| a \right|$=0 }\\
      \text{lev}(\text{tail}(a),\text{tail}(b))  & \text{if $a[0]=b[0]$}\\
      1+\min
      \left\{
      \begin{array}{l}
      \text{lev}(\text{tail}(a),b)\\
      \text{lev}(a,\text{tail}(b))\\
      \text{lev}(\text{tail}(a),\text{tail}(b))\\
      \end{array}
      \right.
      & \text{otherwise}
\end{array}
\right. \]
where for a string x:
\begin{itemize}
\item $\lvert x \rvert$ is defined as the length of the string x, 
\item $\text{tail}(x)$ is defined as the string created from x by removing the first character, 
\item x[n] is the n-th character of x with a zero-based counting.
\end{itemize}
Now that we have a metric that measures distances between strings, we define the distance between two clusters of strings as the average of the pairwise distances. 
Concretely, if we have two clusters $A$ and $B$, each representing a set of strings, we define the distance between $A$ and $B$ as 
\[
l(A,B) =
\frac{1}{ \left| A \right| \cdot \left| B \right|} \sum_{a \in A}^{}\sum_{b \in B}^{} \text{lev}(a,b). 
\]
We can now deploy the agglomerate clustering at threshold $c > 0$ to assign each string to a unique cluster as follows:
We initialize the algorithm considering each string as its own cluster. 
Then we merge two clusters $A_i$ and $A_j$ whose distance $l(A_i, A_j)$ is smaller than $c$ into a single cluster. 
We repeat the previous step iteratively as long as we find at least one pair of clusters with distance $l$ below $c$.

An appropriate clustering threshold $c$ for the roles can be determined by directly looking at the roles that are clustered at each step.
While thresholds of up to 2.5 cluster alternative spellings and spelling mistakes, 
the first wrongly clustered roles are `appellee' and `appellant' at a distance of 3.
These common roles should be distinguished by the clustering mechanism; therefore, a clustering threshold between 2.5 and 3 is chosen, namely 2.7.

Due to the large number of law firm names, determining an appropriate clustering threshold requires a different approach:
We extract a list of law firm names from three major law firm rankings (see \ref{SI:Law_Firm_Rankings}).
We call this list `list 1' and assume that it is free of spelling errors. 
We call the list containing the law firm names extracted from the CAP data `list 2'.
We expect many spelling mistakes in list 2 but not in list 1. 
Further, we expect spelling variations to be `minor in the sense that the Levenshtein distance between two variations of one law firm name is smaller than some threshold.
Hence, if we apply the agglomerative clustering to both list 1 and list 2, we expect that:
\begin{itemize}
    \item For low clustering thresholds the relative decrease in the number of law firm names in list 2 will be steeper than in list 1 because for low thresholds spelling variations are grouped together.
    \item For high clustering thresholds, we expect the relative decrease to be comparable in the two lists.
\end{itemize}
For a given clustering threshold $c$, we run the agglomerative clustering both on the list of official rankings (list 1) and our own list (list 2).
Then, we gradually vary $c$ from 0 to 5, and for each value we compare the number of law firm names in each list.
In list 1, increasing the clustering threshold from 0.1 to 1.1, 2.1, 3.1 and 4.1 reduces the number of law firm names by 1\%, 1\%, 3\%, and 5\%, respectively.
In list 2, increasing the clustering threshold from 0.1 to 1.1, 2.1, 3.1 and 4.1 reduces the number of law firm names by 8\%, 6\%, 5\%, and 6\%, respectively.
This can be explained as follows:
\begin{itemize}
    \item Names grouped with a clustering threshold smaller than 2.1 differ mainly due to spelling variations.
    \item With clustering thresholds higher than 3.1, genuinely different law firm names are predominantly grouped together.
\end{itemize}
Therefore, a clustering threshold between 2.1 and 3.1, namely 2.7, is chosen for the name of the law firms.

Finally, the most frequent representative of every cluster is assigned to every element of the cluster.
For example, in \{`putman $\&$ putman', `put-man and put-man', `putman $<\&$ putman'\} with frequencies (100, 5, 1) `putman $\&$ putman' would replace all three strings.

After executing the above step, we are left with 67,619 cases across 64,656 law firms. 
We refer to Supplementary Table \ref{tab:data_summary} for an overview.

\section{Sub-sampling Pairwise Interactions via $Q$-Factor}
\label{SI:Q-factor}

As described in the \textit{Data} section of the main article and summarized in Supplementary Table \ref{tab:data_summary}, 
our training dataset consists of 167,439 pairwise interactions among law firms. 
These interactions form a network where nodes represent law firms and edges represent cases in which two firms opposed each other, one representing the plaintiff and the other the defendant.  
This \textit{interaction network} is highly heterogeneous, with some firms appearing in numerous cases, while others have only a handful of observed interactions.  
To systematically analyze the structure of this network and ensure statistical robustness, we introduce the \textit{$Q$-factor}, defined as $Q = N/K$,  
where $N$ is the number of observed pairwise interactions and $K$ is the number of unique law firms in the subset.  
Thus, the $Q$-factor represents the average number of interactions per law firm.  

To refine the dataset and focus on firms with sufficient litigation experience, we apply a \textit{trimming} procedure that enforces a minimum $Q$-factor.  
This is achieved by iteratively removing the least-connected law firms ----those with the fewest interactions ---- until the desired $Q$ is reached.  
As a result, this process extracts a \textit{denser core} of the interaction network, where firms have a larger number of observed cases against similarly experienced opponents.    
This approach ensures that retained law firms have a statistically significant number of interactions, improving the reliability of inferred rankings
(see \ref{SI:synth_data} for additional results on data stability across different $Q$ factors).  

\begin{figure}[!htb]
    \centering
    \includegraphics[width=0.6\linewidth]{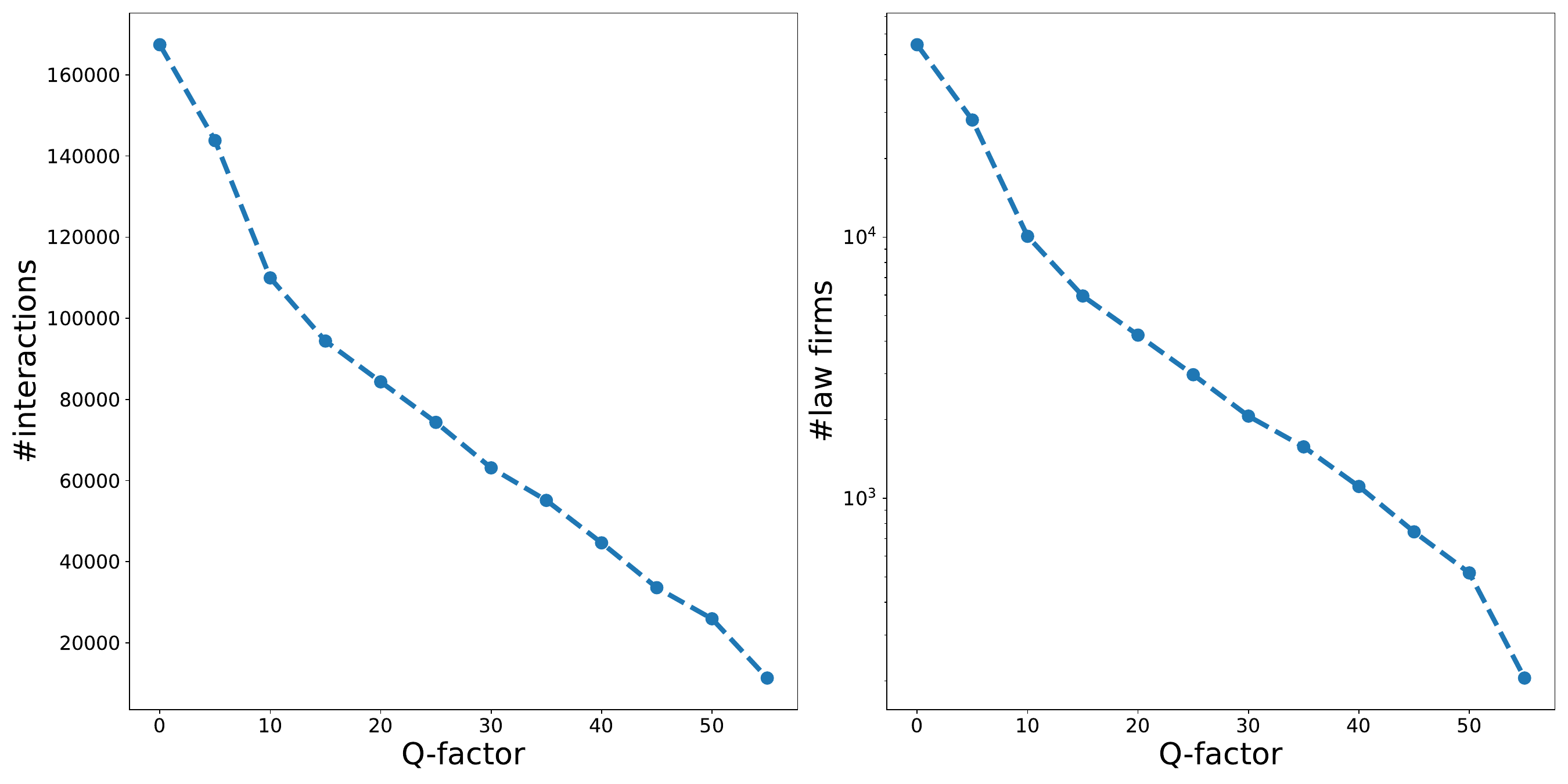}
    \caption{ 
            Decreasing number of interactions (left) and entities (right) as the $Q$-factor is increased.
            }
    \label{fig:Q_basic_plots}
\end{figure}

We now provide some summary statistics of our data as a function of $Q$.
Supplementary Figure \ref{fig:Q_basic_plots} (left) shows a linear decrease in the number of interactions as $Q$ increases, 
while Supplementary Figure \ref{fig:Q_basic_plots} (right) shows an exponential decrease in the number of firms. 
In Supplementary Figure \ref{fig:Q_descriptive_plots} (left), we observe that more recent interactions are more common for a higher $Q$-factor.
As expected by construction of the $Q$-factor, 
its increase leads to an increased lower bound for the degrees of firms in the sample (Supplementary Figure \ref{fig:Q_descriptive_plots} (middle)).
A higher $Q$-factor is also associated with firms participating across more different case types which can be seen in Supplementary Figure \ref{fig:Q_descriptive_plots} (right).

\begin{figure}[!htb]
    \centering
    \includegraphics[width=\linewidth]{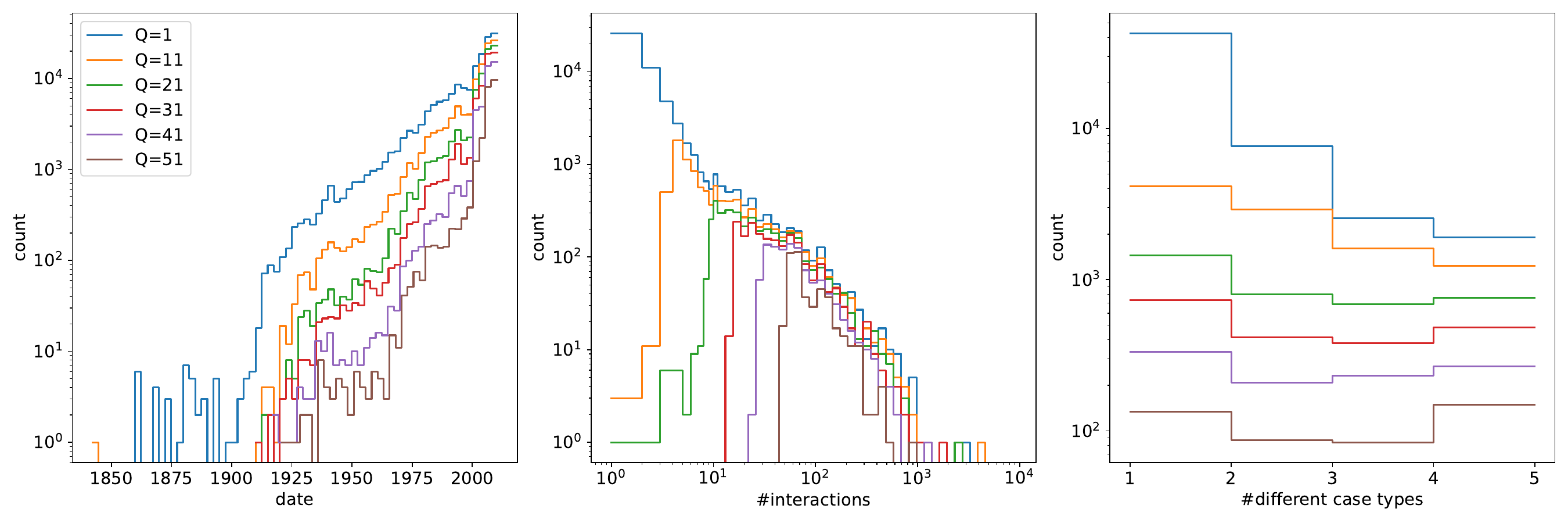}
    \caption{
            We show the distribution of dates (left), number of interactions (middle) and different case types (right) for different values of $Q$.
            }
    \label{fig:Q_descriptive_plots}
\end{figure}

\section{Case Statistics}
\label{SI:Case_Statistics}

\begin{figure}[!htb]
    \centering
    \includegraphics[width=\linewidth]{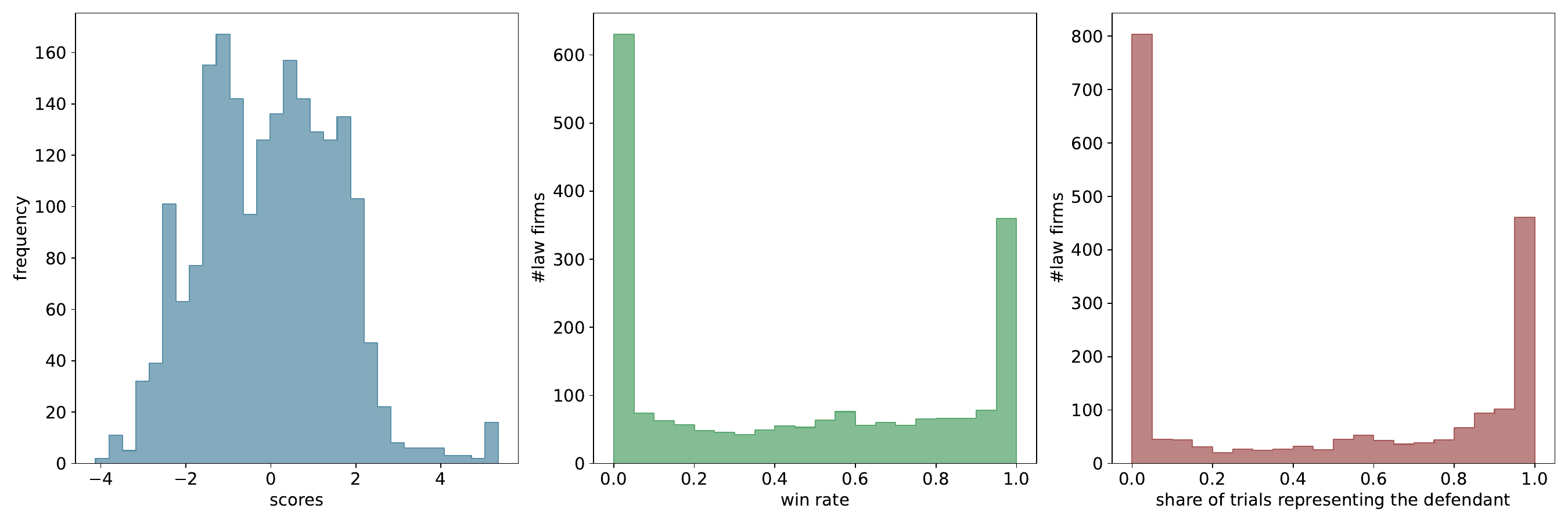}
    \caption{
            (left) Distribution of fitted scores. 
            (middle) Distribution of empirically observed win rates. 
            (right) Distribution of fraction of share of trials as representative of the defendant across law firms.
            }
    \label{fig:fitted_scores_histo}
\end{figure}

For the intermediate $Q$-factor of 30, 
Supplementary Figure \ref{fig:fitted_scores_histo} (left) displays the distribution of the fitted scores. 
The scores are distributed around 0 by construction of AHPI.
Supplementary Figure \ref{fig:fitted_scores_histo} (middle) shows the bimodal distribution of empirical win rates.
This suggests that a large fraction of firms either mostly lose or mostly win.
Supplementary Figure \ref{fig:fitted_scores_histo} (right) shows the distribution of the shares of trials representing the defendant per law firm. 
Its bimodality suggests that most law firms either specialize on the defendant or the plaintiff side. 

\clearpage
\section{Fitting AHPI Algorithm on Synthetic Data}
\label{SI:synth_data}

We test the fitting accuracy of the AHPI algorithm by generating synthetic data with known ground truth. 
Concretely, we generate $N$ interactions among $K$ entities, and each interaction is of one of $M$ types.
To stay close to our actual data on lawsuits, we choose $N, K$ and $M$ given by our empirical data on lawsuits. 
Figures 2 and 3 of the main article have been obtained for a $Q$-factor of $30$ which results in $63,115$ interactions across $2,064$ firms.
We therefore set $N=63,115, K=2,064$ and $M=5$. 
To generate synthetic data, we proceed as follows:
\begin{enumerate}
    \item 
    We randomly assign to each entity a score drawn at random from our empirically fitted $K$ scores. 
    \item 
    We generate an interaction by randomly selecting two of the $K$ entities.
    Similarly, we randomly sample the interaction type $m \in \{1, \ldots, M\}$ in proportion to their empirical occurrence (Table 1 in the main article).
    \item 
    For each interaction, a favored individual is chosen proportionally to the probability \eqref{eq:favoured_probability}. 
    As bias $\epsilon_m$, we use our empirically fitted value $\hat{\epsilon}_m$.
    \item 
    The winner of the interaction is chosen on the basis of \eqref{eq:winner_probability} with $q_m$ replaced by our
empirically fitted value $\hat{q}_m$.
\end{enumerate}
We generate 80 different sets of synthetic data as described in steps $1-4$ above and fit the AHPI algorithm to each. 
We then calculate the average value of each fitted parameter across the 80 sets and use the standard deviations as error bars. 
The result is depicted in Supplementary Table \ref{tab:fitted_synthetic}.
We notice that the true scores and the fitted scores are, on average, 81\% correlated.
In line with our empirical fits, the valence probabilities are consistently large. 
The values for $\epsilon$ are close to the real values, although with slight biases. 
These results suggest that our fitted scores capture the relevant tendencies, but there is room for improvement.
We leave this for future research, and refer to refs. \cite{Newman2022,Jerdee2023} for more discussions on this topic. 

\begin{table}[ht]
\centering

\begin{tabular}{l|cc}
\hline
 & \textbf{actual data} & \textbf{synthetic data} \\
\hline\hline
scores Kendall's $\tau$ & $-$ & 0.85 $\pm$ 0.09 \\
\hline
$\epsilon_{\text{civil rights}}$ & 2.03 & 1.77 $\pm$ 0.21 \\
\hline
$\epsilon_{\text{contract}}$ & 1.66 & 1.46 $\pm$ 0.17 \\
\hline
$\epsilon_{\text{torts}}$ & 0.32 & 0.29 $\pm$ 0.04 \\
\hline
$\epsilon_{\text{labor}}$ & 2.00 & 1.76 $\pm$ 0.20 \\
\hline
$\epsilon_{\text{other}}$ & 1.90 & 1.57 $\pm$ 1.36 \\
\hline\hline
$q_{\text{civil rights}}$ & 0.86 & 0.88 $\pm$ 0.04 \\
\hline
$q_{\text{contract}}$ & 0.96 & 0.97 $\pm$ 0.05 \\
\hline
$q_{\text{torts}}$ & 1.00 & 0.99 $\pm$ 0.06 \\
\hline
$q_{\text{labor}}$ & 0.96 & 0.97 $\pm$ 0.05 \\
\hline
$q_{\text{other}}$ & 1.00 & 0.98 $\pm$ 0.10 \\
\hline\hline
\end{tabular}

\caption{
The left column shows the fitted model parameters on empirical data with $Q=30$.
The right column shows average and standard deviation of fits on synthetic data. 
In particular, the top row shows the correlation between fitted and true values. 
}
\label{tab:fitted_synthetic}
\end{table}

\begin{figure}[!htb]
    \centering
    \includegraphics[width=\linewidth]{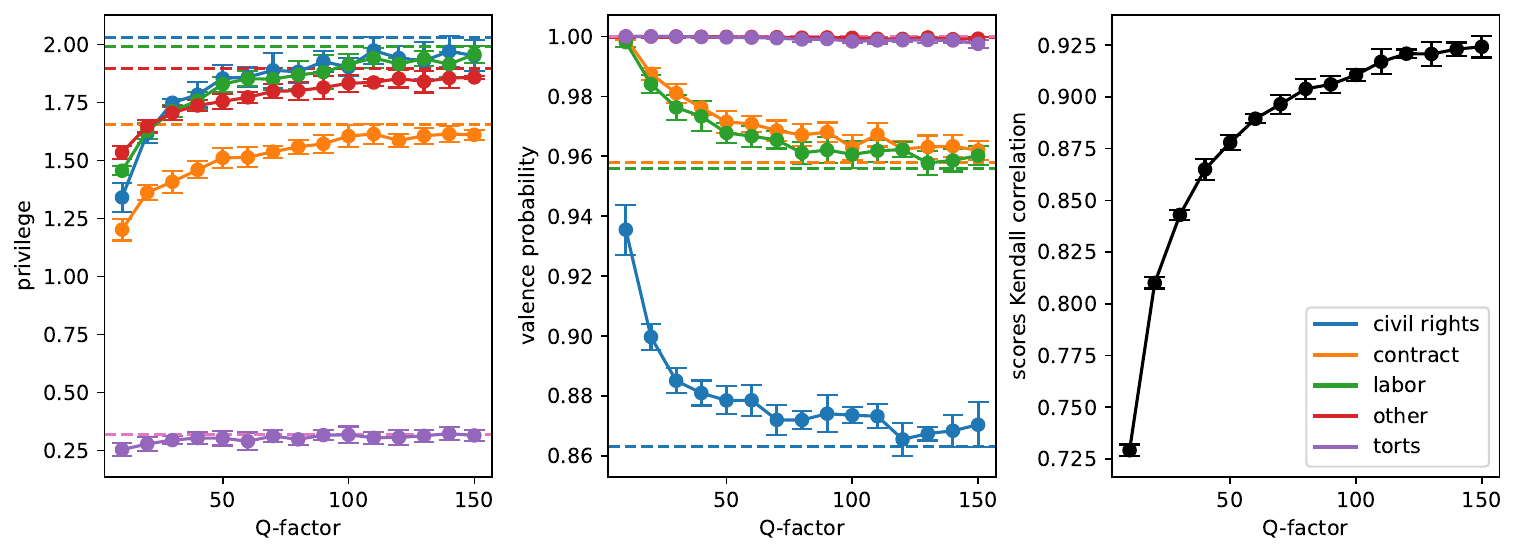}
    \caption{
            AHPI fitting accuracy as a function of the $Q$-factor. 
            (left)
            Fitted biases $\epsilon_m$. 
            Dashed lines indicate ground truth values. 
            (middle)
            Fitted valence probabilities $q_m$. 
            Similar to the fits of $\epsilon_m$, we notice slight biases. 
            (right)
            Kendall correlation between ground truth scores and fitted scores.
            }
    \label{fig:Q_fact_fitting}
\end{figure}

A central parameter used to characterize data is the ratio of the number of interactions $R$ to the number of entities $N$, which we call the $Q$-factor (see above). 
We expect that as the $Q$-factor increases, the AHPI algorithm becomes more accurate.
To test this, we iteratively increase the $Q$-factor by removing entities with the lowest number of interactions, 
and repeat the above described fitting procedure on synthetic data for a sequence of $Q$-factors. 
The outcome is depicted in Supplementary Figure \ref{fig:Q_fact_fitting} and in line with our expectation: 
The higher the $Q$-factor, the closer the fitted values are to the ground truth. 
However, we notice that slight biases exist even when $Q$ becomes large. 
We leave a more detailed examination of this for future research, and refer to refs. \cite{Newman2022,Jerdee2023} for more discussions on this topic.

\clearpage
\section{Legal Case Outcome Predictions for Various $Q$-Factors}
\label{SI:Q_factor_predictions}

Recall from above that the $Q$-factor is defined as the average number of interactions per law firm in a given dataset.
In Figure 3 of the main paper we have shown prediction accuracies on a subset where $Q=30$. 
In Supplementary Figure \ref{fig:various_Q_pred}, we show that qualitatively similar results for a wide range of $Q$-factors. 

\begin{figure}[ht!]
    \centering
    \begin{minipage}{0.45\textwidth}
        \includegraphics[width=\linewidth]{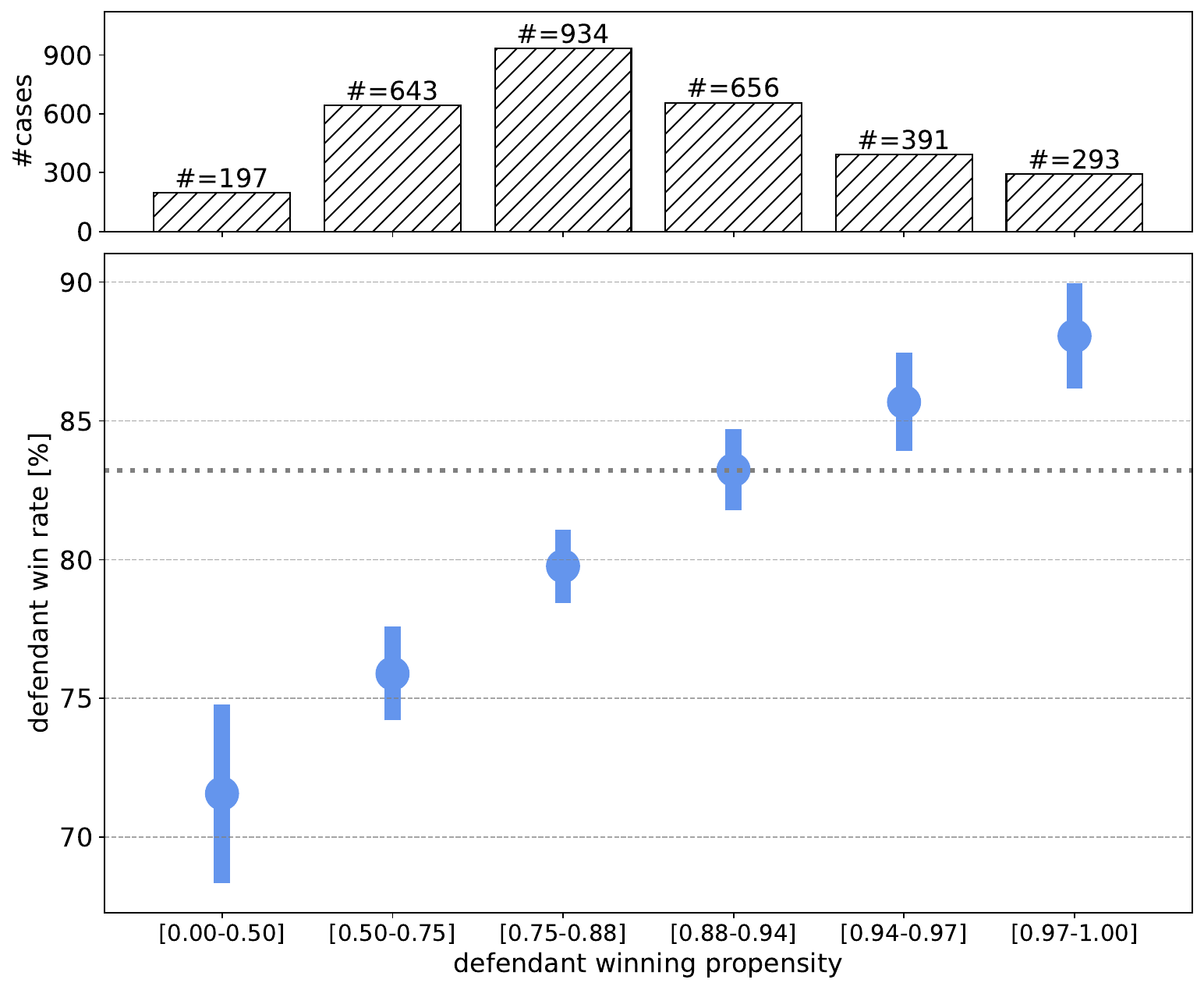}
        \caption*{(1) $Q = 20$}
    \end{minipage}\hfill
    \begin{minipage}{0.45\textwidth}
        \includegraphics[width=\linewidth]{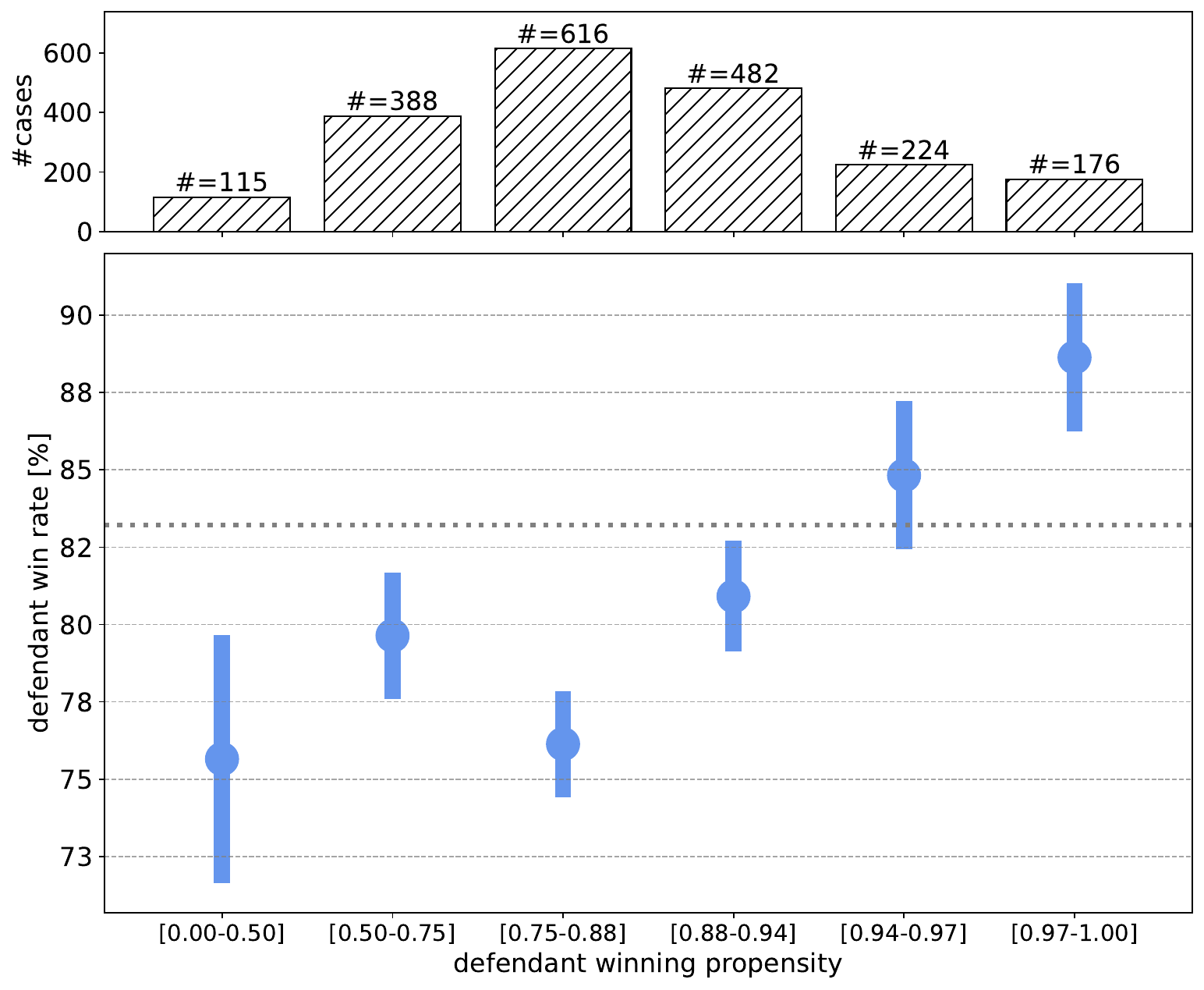}
        \caption*{(2) $Q = 25$}
    \end{minipage}

    \begin{minipage}{0.45\textwidth}
        \includegraphics[width=\linewidth]{./figures/figures_q/predictions_Q_30.pdf}
        \caption*{(3) $Q = 30$}
    \end{minipage}\hfill
    \begin{minipage}{0.45\textwidth}
        \includegraphics[width=\linewidth]{./figures/figures_q/predictions_Q_35.pdf}
        \caption*{(4) $Q = 35$}
    \end{minipage}
\end{figure}
\begin{figure}

    \begin{minipage}{0.45\textwidth}
        \includegraphics[width=\linewidth]{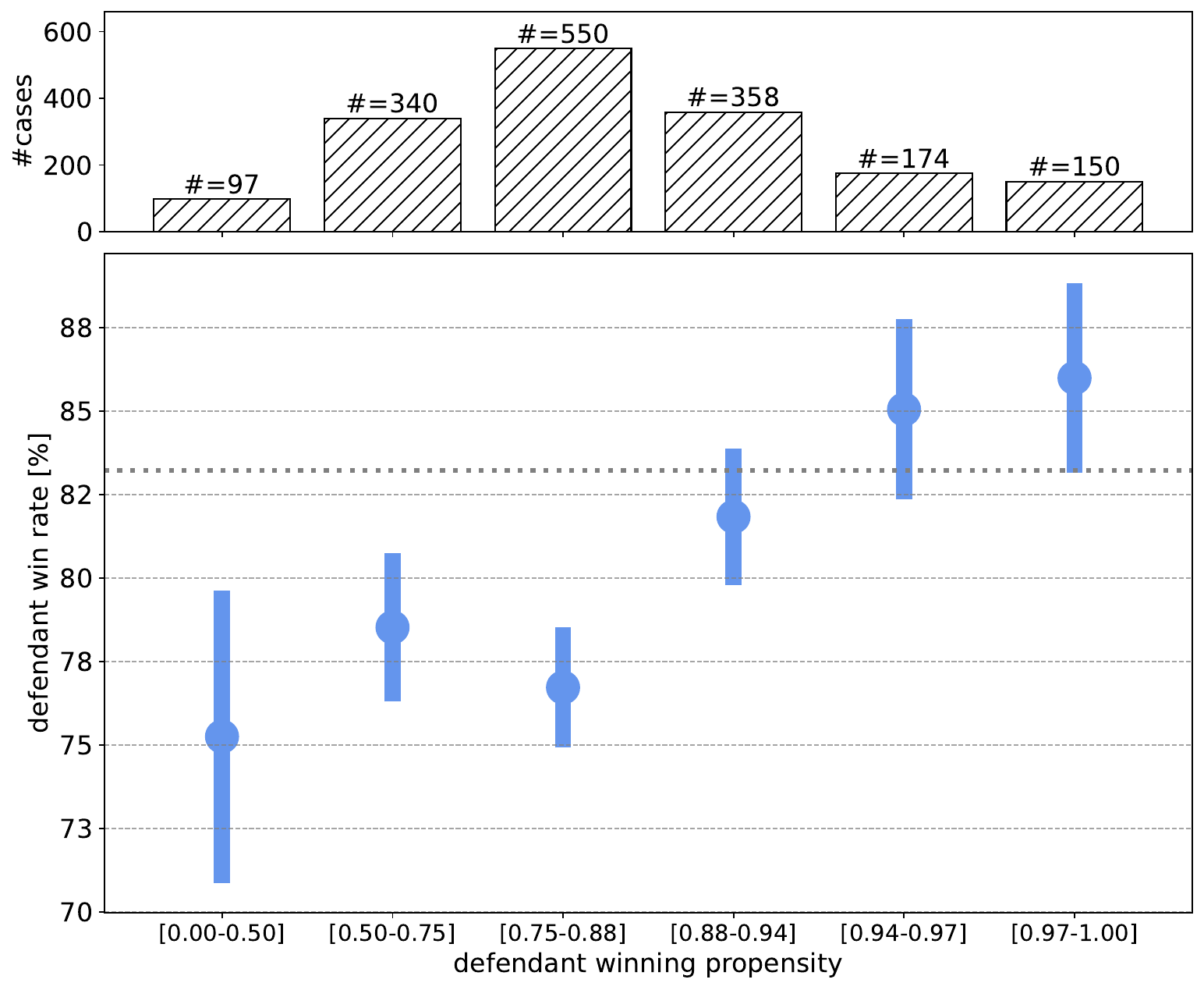}
        \caption*{(5) $Q = 40$}
    \end{minipage}\hfill
    \begin{minipage}{0.45\textwidth}
        \includegraphics[width=\linewidth]{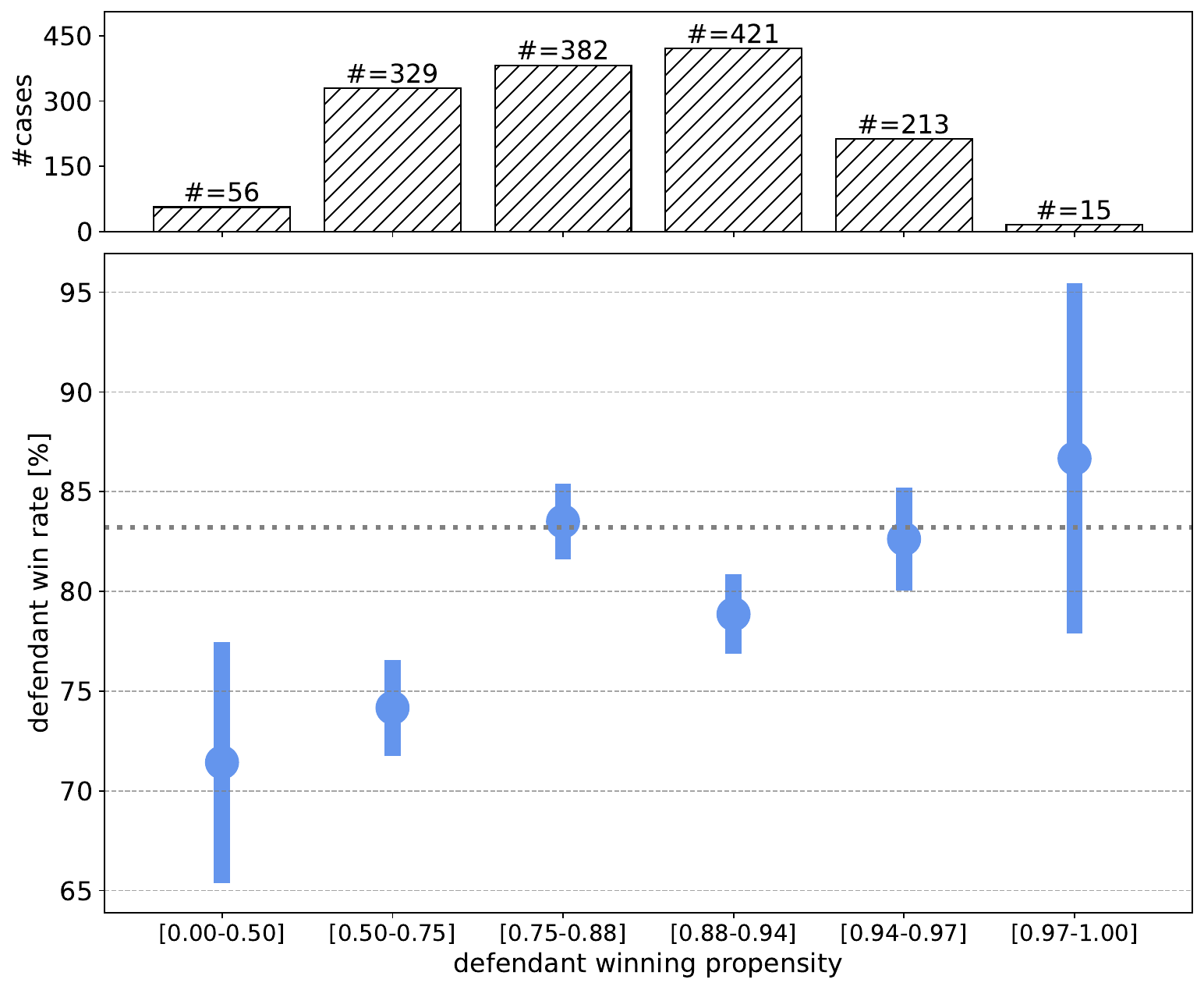}
        \caption*{(6) $Q = 45$}
    \end{minipage}
    \begin{minipage}{0.45\textwidth}
        \includegraphics[width=\linewidth]{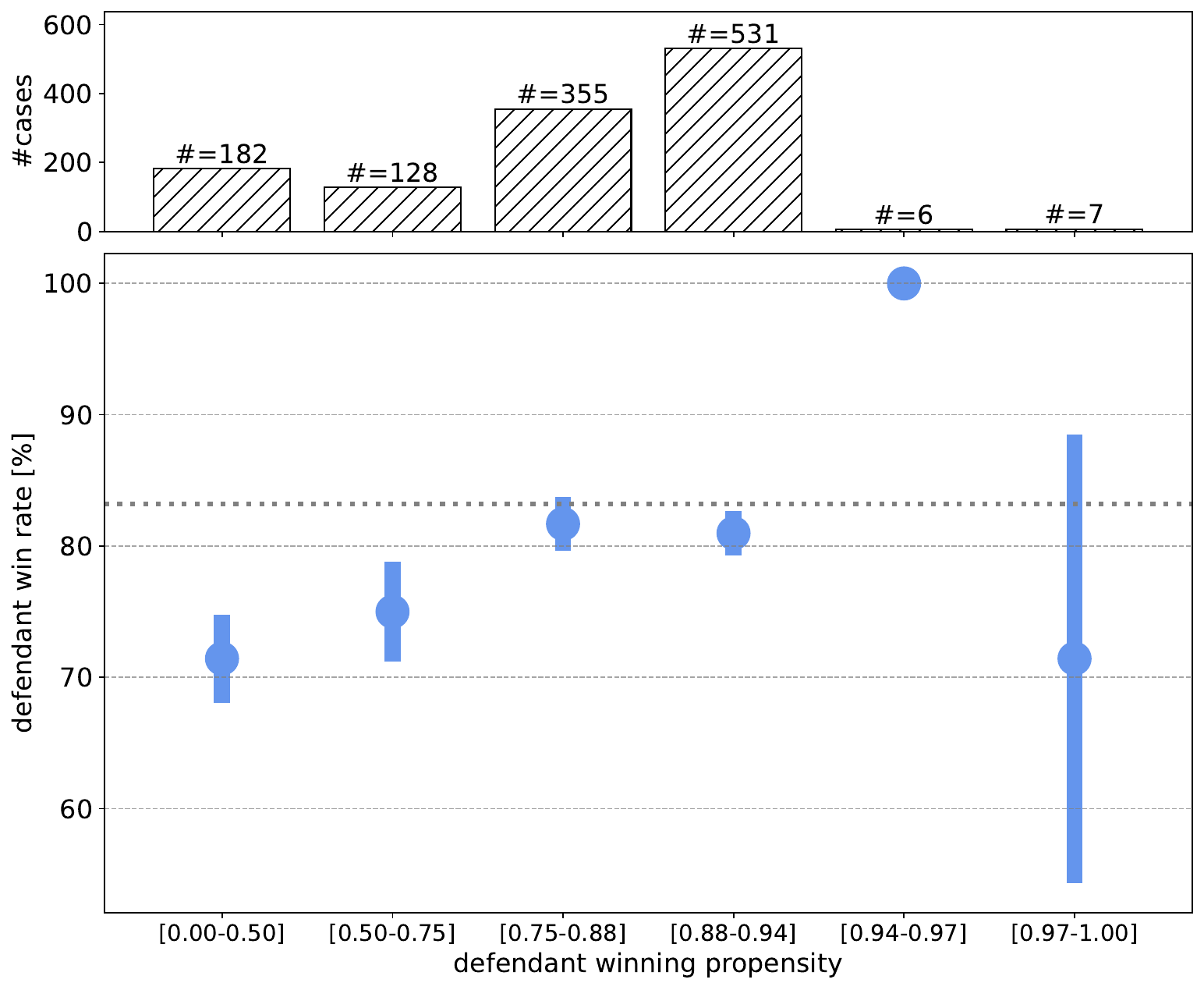}
        \caption*{(5) $Q = 50$}
    \end{minipage}\hfill
    \begin{minipage}{0.45\textwidth}
        \includegraphics[width=\linewidth]{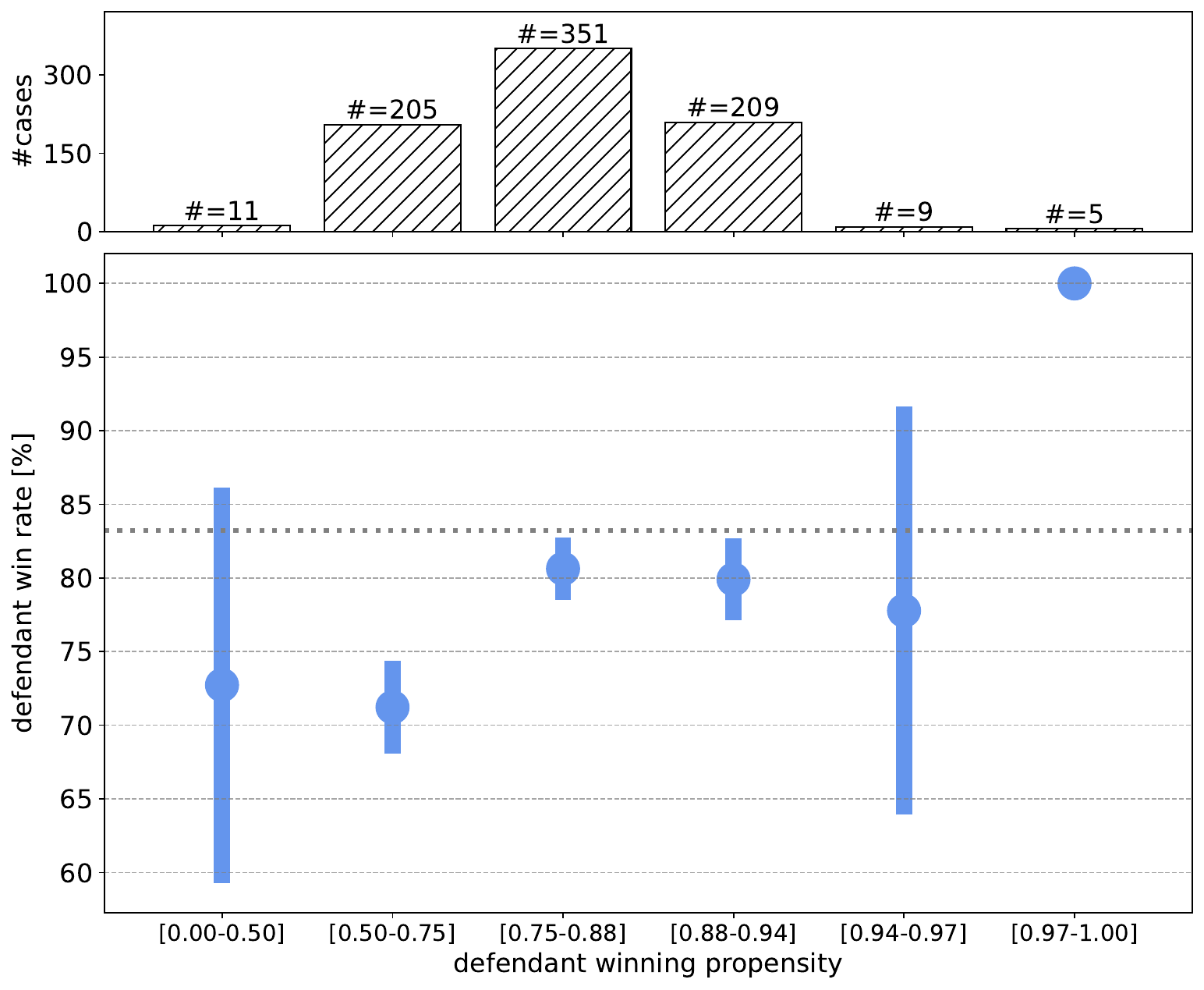}
        \caption*{(6) $Q = 55$}
    \end{minipage}
    \caption{
    Same as Fig 3 in the main, but for different $Q$-factors.}
    \label{fig:various_Q_pred}
\end{figure}

\clearpage
\section{Use of Sigmoid Function in Ranking Algorithm}
\label{SI:sigmoid_usage}

Our ranking algorithm is based on the Bradley-Terry model, which is commonly used to study heterogeneous pairwise interactions~\cite{Newman2022,Jerdee2023}.  
In this Appendix, we explain why the sigmoid function is used to map score differences onto predicted winning probabilities and discuss the rationale behind this choice.

For two individuals $i$ and $j$ with scores $S_i$ and $S_j$, respectively, the function $f(s)$ that maps the score difference $S \equiv S_i - S_j$ to a probability should satisfy the following axioms:

\renewcommand{\labelenumi}{\Alph{enumi}.}
\begin{enumerate}
    \item $f(S)$ is increasing in $S$, ensuring that a higher score for individual $i$ results in a higher probability of $i$ winning.
    \item $\lim_{S \to -\infty} f(S) = 0$ and $\lim_{S \to \infty} f(S) = 1$, so that an infinitely strong (weak) player always wins (loses).
    \item $f(-S) = 1 - f(S)$, ensuring that the sum of the probabilities of $i$ and $j$ winning is equal to 1.
\end{enumerate}

These axioms define a set of candidate functions that could be used. Among these, the sigmoid function is a natural choice for two primary reasons:
First, the sigmoid function, $f(S) = \frac{1}{1 + e^{-S}}$, is analytically simple, computationally efficient, and has a straightforward inverse and derivative.
Second, the sigmoid function is directly linked to Bayesian inference in binary outcomes, where a result is either `win' or `loss' (e.g., `i wins' versus `j wins'). 
Specifically, we can interpret the score difference $S$ as the logarithm of the odds ratio between the two players, which is a standard approach in logistic regression and other probabilistic models.

In a Bayesian framework, for the binary outcome setting ``$i$ wins'' versus ``$j$ wins,'' given some information $X$ (i.e. the players' respective scores) the following holds:
\[
p(\text{$i$ wins} \mid X) = \frac{1}{1 + \text{Odds}^{-1}} = \frac{1}{1 + \exp\left(-\log(\text{Odds}) \right)},
\]
where the odds ratio is given by:
\[
\text{Odds} \equiv \frac{p(X \mid \text{$i$ wins})\,p(\text{$i$ wins})}{p(X \mid \text{$j$ wins})\,p(\text{$j$ wins})}.
\]
Identifying the score difference $S$ with $\log(\text{Odds})$ recovers the sigmoid function. 
Thus, using the sigmoid to map score differences directly aligns with a Bayesian framework for binary outcome probabilities, making it both a theoretically grounded and computationally efficient choice.
This interpretation is intuitive because it links the score difference between two players to their relative winning odds, providing a clear probabilistic interpretation of the model output.

\clearpage
\section{Public Law Firm Rankings}
\label{SI:Law_Firm_Rankings}

A natural source to contextualize our approach is to compare our ranking with pre-existing ones. 
We have drawn on three common ranking: 

\begin{itemize}

    \item 
    The \textit{Vault Law 100} ranking is determined by averaging the prestige scores from anonymous surveys completed by law firm associates across the U.S., focusing solely on the perceived prestige of working for these firms rather than their financial or operational metrics.
    The ranking we used was based on surveys conducted from January to March 2013 to match our train-test split cutoff, which coincides with that year.
    \footnote{Retrieved from https://firsthand.co/best-companies-to-work-for/law/top-100- law-firms-rankings in February 2023.}
    The \textit{Vault100} is the most prevalent law firm ranking and among the three presented rankings the only one providing actual scores as opposed to just rankings. 
    We visualize our ranking scores versus the \textit{Valut100} ranking scores in the Supplementary Figure \ref{fig:scatter_vault100}.
    
    \item 
    The \textit{Embroker Law 300} Award ranking distinguishes itself by evaluating law firms on innovative criteria such as the adoption of advanced courtroom technology, initiatives for environmental sustainability, progressive employment practices, and efforts to ensure equal access to justice, rather than traditional metrics like revenue and firm size.
    We use the 2022 ranking.\footnote{Retrieved from https://www.embroker.com/blog/top-300- law-firms/ in February 2023.}
    
    \item 
    \textit{ALM's Global 200} ranking is calculated based on the analysis of key financial metrics such as revenue, profits, and headcount growth for each of the world's 200 largest law firms.
    We use the 2022 ranking.\footnote{Retrieved from https://www.law.com/law-firms/ in February 2023.}
\end{itemize}

\begin{figure}[!htb]
    \centering
    \includegraphics[width=0.7\linewidth]{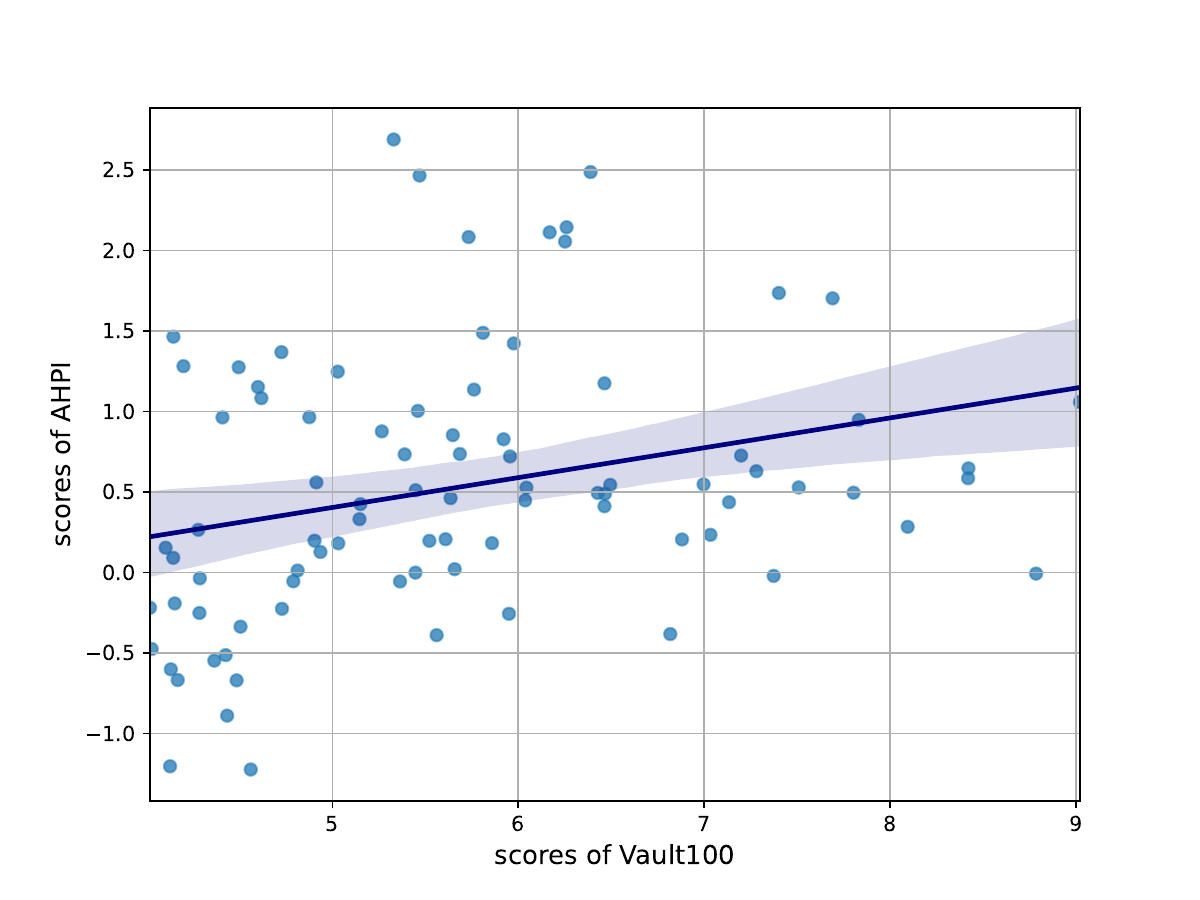}
    \caption{
            Scatter plot visualizing \textit{Vault Law 100} scores from 2022 vs. the scores fitted by our AHPI algorithm. 
            No significant correlation is found.
            }
    \label{fig:scatter_vault100}
\end{figure}

\end{document}